\documentclass[prd,aps,floats,onecolumn]{revtex4}
\usepackage{amsmath}
\usepackage{amssymb}

\begin{document}

\title{Nonlinear Realization of the Local Conform-Affine Symmetry Group for
Gravity in the Composite Fiber Bundle Formalism}

\author{S.A. Ali${}^{\flat}$, S. Capozziello${}^{\sharp}$}
\affiliation{${}^{\flat}$Department of Physics, University at
Albany-SUNY, 1400 Washington Avenue, Albany, NY, 12222, USA
\\ ${}^{\sharp}$ Dipartimento di Scienze Fisiche, Universit\`a di Napoli "Federico II" and INFN Sez. di Napoli,
Compl. Univ. Monte S. Angelo, Ed.N, Via Cinthia, I-80126 Napoli,
Italy}

\begin{abstract}
A gauge theory of gravity based on a nonlinear realization (NLR)
of the local Conform-Affine (CA) group of symmetry transformations
is presented. The coframe fields and gauge connections of the
theory are obtained. The tetrads and Lorentz group metric are used
to induce a spacetime metric. The inhomogenously transforming
(under the Lorentz group) connection coefficients serve as
gravitational gauge potentials used to define covariant
derivatives accommodating minimal coupling of matter and gauge
fields. On the other hand, the tensor valued connection forms
serve as auxillary dynamical fields associated with the dilation,
special conformal and deformational (shear) degrees of freedom
inherent in the bundle manifold. The bundle curvature of the
theory is determined. Boundary topological invariants are
constructed. They serve as a prototype (source free) gravitational
Lagrangian. The Bianchi identities, covariant field equations and
gauge currents are obtained.\\

Key Words: gauge symmetry, conform-affine Lie algebra, gravity,
fiber bundle formalism.

\end{abstract}

\maketitle

\section{Introduction}

Quantum theory and relativity theory are two fundamental theories in modern
physics. The so-called standard model is currently the most successful
relativistic quantum theory in particle physics. It is a non-Abelian gauge
theory (Yang-Mills theory) associated with the internal symmetry group $%
SU(3)\times SU(2)\times U(1)$, in which the $SU(3)$ color symmetry for the
strong force in quantum chromodynamics is treated as exact whereas the $%
SU(2)\times U(1)$ symmetry responsible for generating the electro-weak gauge
fields is spontaneously broken. So far as we know, there are four
fundamental forces in Nature; namely, electromagnetic force, weak force,
strong force and gravitational force. The standard model covers the first
three, but not the gravitational interaction. In general relativity, the
geometrized gravitational field is described by the metric tensor $g_{\mu
\nu }$ of pseudo-Riemannian spacetime, and the field equations that the
metric tensor satisfies are nonlinear. This nonlinearity is indeed a source
of difficulty in quantization of general relativity. Since the successful
standard model in particle physics is a gauge theory in which all the fields
mediating the interactions are represented by gauge potentials, a question
arises as to why the fields mediating the gravitational interaction are
different from those of other fundamental forces. It is reasonable to expect
that there may be a gauge theory in which the gravitational fields stand on
the same footing as those of other fields. This expectation has prompted a
re-examination of general relativity from the gauge theoretical point of
view.

While the gauge groups involved in the standard model are all internal
symmetry groups, the gauge groups in general relativity must be associated
with external spacetime symmetries. Therefore, the gauge theory of gravity
will not be a usual Yang-Mills theory. It must be one in which gauge objects
are not only the gauge potentials but also tetrads that relate the symmetry
group to the external spacetime. For this reason we have to consider a more
complex nonlinear gauge theory. In general relativity, Einstein took the
spacetime metric as the basic variable representing gravity, whereas
Ashtekar employed the tetrad fields and the connection forms as the
fundamental variables. We also consider the tetrads and the connection forms
as the fundamental fields.

R. Utiyama (1956) was the first to suggest that gravitation may be viewed as
a gauge theory \cite{Utiyama} in analogy to the Yang-Mills \cite{YangMills}
theory (1954). He identified the gauge potential due to the Lorentz group
with the symmetric connection of Riemann geometry, and constructed
Einstein's general relativity as a gauge theory of the Lorentz group $SO(3$,
$1)$ with the help of tetrad fields introduced in an \textit{ad hoc} manner.
Although the tetrads were necessary components of the theory to relate the
Lorentz group adopted as an internal gauge group to the external spacetime,
they were not introduced as gauge fields. In 1961, T.W.B. Kibble \cite%
{Kibble} constructed a gauge theory based on the Poincar\'{e} group $P(3$, $%
1)=T(3$, $1)\rtimes SO(3$, $1)$ ($\rtimes $ represents the semi-direct
product) which resulted in the Einstein-Cartan theory characterized by
curvature and torsion. The translation group $T(3$, $1)$ is considered
responsible for generating the tetrads as gauge fields. Cartan \cite{Cartan}
generalized the Riemann geometry to include torsion in addition to
curvature. The torsion (tensor) arises from an asymmetric connection. D.W.
Sciama \cite{Sciama}, and others (R. Fikelstein \cite{Finkelstein}, Hehl
\cite{Hehl1, Hehl2}) pointed out that intrinsic spin may be the source of
torsion of the underlying spacetime manifold.

Since the form and role of the tetrad fields are very different from those
of gauge potentials, it has been thought that even Kibble's attempt is not
satisfactory as a full gauge theory. There have been a number of gauge
theories of gravitation based on a variety of Lie groups \cite{Hehl1, Hehl2,
Mansouri1, Mansouri2, Chang, Grignani, MAG}. It was argued that a gauge
theory of gravitation corresponding to general relativity can be constructed
with the translation group alone in the so-called teleparallel scheme.
Inomata \textit{et al.} \cite{Inomata} proposed that Kibble's gauge theory
could be obtained in a manner closer to the Yang-Mills approach by
considering the de Sitter group $SO(4$, $1)$ which is reducible to the
Poincar\'{e} group by group-contraction. Unlike the Poincar\'{e} group, the
de Sitter group is homogeneous and the associated gauge fields are all of
gauge potential type. By the Wigner-In\"{o}nu group contraction procedure,
one of five vector potentials reduces to the tetrad.

It is common to use the fiber-bundle formulation by which gauge theories can
be constructed on the basis of any Lie group. Recent work by Hehl \textit{et
al.} \cite{MAG} on the so-called Metric Affine Gravity (MAG) theory adopted
as a gauge group the affine group $A(4$, $\mathbf{%
%TCIMACRO{\U{211d} }%
%BeginExpansion
\mathbb{R}
%EndExpansion
})=T(4)\rtimes GL(4$, $\mathbf{%
%TCIMACRO{\U{211d} }%
%BeginExpansion
\mathbb{R}
%EndExpansion
})$ which was realized linearly. The tetrad was identified with the
nonlinearly realized translational part of the affine connection on the
tangent bundle. In MAG theory, the Lagrangian is quadratic in both curvature
and torsion in contrast to the Einstein-Hilbert Lagrangian in general
relativity which is linear in the scalar curvature. The theory has the
Einstein limit on one hand and leads to the Newtonian inverse distance
potential plus the linear confinement potential in the weak field
approximation on the other. As we have seen above, there are many attempts
to formulate gravitation as a gauge theory. Currently no theory has been
uniquely accepted as the gauge theory of gravitation.

The nonlinear approach to group realizations was originally introduced by S.
Coleman, J. Wess and B. Zumino \cite{CCWZ1, CCWZ2} in the context of
internal symmetry groups (1969). It was later extended to the case of
spacetime symmetries by Isham, Salam, and Strathdee \cite{Isham, Salam}
considering the nonlinear action of $GL(4$, $\mathbf{%
%TCIMACRO{\U{211d} }%
%BeginExpansion
\mathbb{R}
%EndExpansion
})$ mod the Lorentz subgroup. In 1974, Borisov, Ivanov and Ogievetsky \cite%
{BorisovOgievetskii, IvanovOgievetskii} considered the simultaneous
nonlinear realization (NLR) of the affine and conformal groups. They showed
that general relativity can be viewed as a consequence of spontaneous
breakdown of the affine symmetry in much the same manner that chiral
dynamics in quantum chromodynamics is a result of spontaneous breakdown of
chiral symmetry. In their model, gravitons are considered as Goldstone
bosons associated with the affine symmetry breaking. In 1978, Chang and
Mansouri \cite{ChangMansouri} used the NLR scheme employing $GL(4$, $\mathbf{%
%TCIMACRO{\U{211d} }%
%BeginExpansion
\mathbb{R}
%EndExpansion
})$ as the principal group. In 1980, Stelle and West \cite{StelleWest}
investigated the NLR induced by the spontaneous breakdown of $SO(3$, $2)$.
In 1982 Ivanov and Niederle considered nonlinear gauge theories of the
Poincar\'{e}, de Sitter, conformal and special conformal groups \cite%
{IvanovNiederle1, IvanovNiederle2}. In 1983, Ivanenko and Sardanashvily \cite%
{IvanenkoSardanashvily} considered gravity to be a spontaneously broken $%
GL(4 $, $\mathbf{%
%TCIMACRO{\U{211d} }%
%BeginExpansion
\mathbb{R}
%EndExpansion
})$ gauge theory. The tetrads fields arise in their formulation as a result
of the reduction of the structure group of the tangent bundle from the
general linear to Lorentz group. In 1987, Lord and Goswami \cite{Lord1,
Lord2} developed the NLR in the fiber bundle formalism based on the bundle
structure $G\left( G/H\text{, }H\right) $ as suggested by Ne'eman and Regge
\cite{NeemanRegge}. In this approach the quotient space $G/H$ is identified
with physical spacetime. Most recently, in a series of papers, A.
Lopez-Pinto, J. Julve, A. Tiemblo, R. Tresguerres and E. Mielke discussed
nonlinear gauge theories of gravity on the basis of the Poincar\'{e}, affine
and conformal groups \cite{Julve, Lopez-Pinto, TresguerresMielke,
Tresguerres, TiembloTresguerres1, TiembloTresguerres2}. In the present
paper, we consider a modified version of the theories proposed by
Tresguerres and Lopez-Pinto \textit{et al.}

The paper is organized as follows. In Section $2$, mainly following
Tresguerres and Tiemblo \cite{Tresguerres, TiembloTresguerres1}, the
generalized bundle structure of gravity is presented. In Section $3$, a
generalized gauge transformation law enabling the gauging of external
spacetime groups is introduced. Demanding that tetrads be obtained as gauge
fields requires the implementation of a NLR of the CA group. Such a NLR is
carried out over the quotient space $CA(3$, $1)$/$SO(3$, $1)$. In Section $4$%
, the transformations of all coset fields parameterizing this quotient space
is computed. The fundamental vector field operators are computed in Section $%
5$. In Section $6$, the general form of the gauge connections of the theory
along with their transformation laws are obtained. In Section $7$, we
present the explicit structure of the CA connections. The nonlinear
translational connection coefficient (transforming as a $4$-covector under
the Lorentz group) is identified as a coframe field. In Section $8$, the
tetrad components of the coframe are used in conjunction with the Lorentz
group metric to induce a spacetime metric. In Section $9$, the bundle
curvature of the theory together with the variations of its corresponding
field strength components are determined. The Bianchi identities are
obtained in Section $10$. In Section $11$, surface ($3D$) and bulk ($4D$)
topological invariants are constructed. The bulk terms (obtained via
exterior derivation of the surface terms) provide a means of "deriving" a
prototype (source free) gravitational action (after appropriately
distributing Lie star operators). The covariant field equations and gauge
currents are obtained in Section $12$. Our conclusions are presented in
Section $13$.

\subsection{Ordinary Fiber Bundles, Gauge Symmetry and Connection Forms}

The purpose of this section is to briefly review the standard bundle
approach to gauge theories. We verify that the usual gauge potential $\Omega
$ is the pullback of connection 1-form $\omega $ by local sections of the
bundle. Finally, the transformation laws of the $\omega $ and $\Omega $
under the action of the structure group $G$ are deduced.

Modern formulations of gauge field theories are expressible geometrically in
the language of principal fiber bundles. A fiber bundle is a structure $%
\left\langle \mathbb{P}\text{, }M\text{, }\pi \text{; }\mathbb{F}%
\right\rangle $ where $\mathbb{P}$ (the total bundle space) and $M$ (the
base space) are smooth manifolds, $\mathbb{F}$ is the fiber space and the
surjection $\pi $\ (a canonical projection) is a smooth map of $\mathbb{P}$
onto $M$,%
\begin{equation}
\pi :\mathbb{P}\rightarrow M\text{.}
\end{equation}%
The inverse image $\pi ^{-1}$ is diffeomorphic to $\mathbb{F}$%
\begin{equation}
\pi ^{-1}\left( x\right) \equiv \mathbb{F}_{x}\approx \mathbb{F}\text{,}
\end{equation}%
and is called the fiber at $x\in M$. The partitioning $\bigcup\nolimits_{x}%
\pi ^{-1}\left( x\right) =\mathbb{P}$ is referred to as the fibration. Note
that a smooth map is one whose coordinatization is $C^{\infty }$
differentiable; a smooth manifold is a space that can be covered with
coordinate patches in such a manner that a change from one patch to any
overlapping patch is smooth, see A. S. Schwarz \cite{Schwarz}. Fiber bundles
that admit decomposition as a direct product, locally looking like $\mathbb{%
P\approx }M\times \mathbb{F}$, is called trivial. Given a set of open
coverings $\left\{ \mathcal{U}_{i}\right\} $ of $M$ with $x\in \left\{
\mathcal{U}_{i}\right\} \subset M$ satisfying $\bigcup\nolimits_{\alpha }%
\mathcal{U}_{\alpha }=M$, the diffeomorphism map is given by%
\begin{equation}
\chi _{i}:\mathcal{U}_{i}\times _{M}G\rightarrow \pi ^{-1}(\mathcal{U}%
_{i})\in \mathbb{P}\text{,}  \label{diff}
\end{equation}%
($\times _{M}$ represents the fiber product of elements defined over space $%
M $) such that $\pi \left( \chi _{i}\left( x\text{, }g\right) \right) =x$
and $\chi _{i}\left( x\text{, }g\right) =\chi _{i}\left( x\text{, }\left(
id\right) _{G}\right) g=\chi _{i}\left( x\right) g\ \forall x\in \left\{
\mathcal{U}_{i}\right\} $ and $g\in G$. Here, $\left( id\right) _{G}$
represents the identity element of group $G$. In order to obtain the global
bundle structure, the local charts $\chi _{i}$ must be glued together
continuously. Consider two patches $\mathcal{U}_{n}$ and $\mathcal{U}_{m}$
with a non-empty intersection $\mathcal{U}_{n}\cap \mathcal{U}_{m}\neq
\emptyset $. Let $\rho _{nm}$ be the restriction of $\chi _{n}^{-1}$ to $\pi
^{-1}(\mathcal{U}_{n}\cap \mathcal{U}_{m})$ defined by $\rho _{nm}:\pi ^{-1}(%
\mathcal{U}_{n}\cap \mathcal{U}_{m})\rightarrow (\mathcal{U}_{n}\cap
\mathcal{U}_{m})\times _{M}G_{n}$. Similarly let $\rho _{mn}:\pi ^{-1}(%
\mathcal{U}_{m}\cap \mathcal{U}_{n})\rightarrow (\mathcal{U}_{m}\cap
\mathcal{U}_{n})\times _{M}G_{m}$ be the restriction of $\chi _{m}^{-1}$ to $%
\pi ^{-1}(\mathcal{U}_{n}\cap \mathcal{U}_{m})$. The composite
diffeomorphism $\Lambda _{nm}\in G$%
\begin{equation}
\Lambda _{mn}:(\mathcal{U}_{n}\cap \mathcal{U}_{m})\times G_{n}\rightarrow (%
\mathcal{U}_{m}\cap \mathcal{U}_{n})\times _{M}G_{m}\text{,}
\end{equation}%
defined as%
\begin{equation}
\Lambda _{ij}\left( x\right) \equiv \rho _{ji}\circ \rho _{ij}^{-1}=\chi _{i%
\text{, }x}\circ \chi _{j\text{, }x}^{-1}:\mathbb{F}\rightarrow \mathbb{F}
\end{equation}%
constitute the transition function between bundle charts $\rho _{nm}$ and $%
\rho _{mn}$ ($\circ $ represents the group composition operation) where the
diffeomorphism $\chi _{i\text{, }x}:\mathbb{F}\rightarrow \mathbb{F}_{x}$ is
written as $\chi _{i\text{, }x}(g):=\chi _{i}\left( x\text{, }g\right) $ and
satisfies $\chi _{j}\left( x\text{, }g\right) =\chi _{i}\left( x\text{, }%
\Lambda _{ij}\left( x\right) g\right) $. The transition functions $\left\{
\Lambda _{ij}\right\} $ can be interpreted as passive gauge transformations.
They satisfy the identity $\Lambda _{ii}\left( x\right) $, inverse $\Lambda
_{ij}\left( x\right) =\Lambda _{ji}^{-1}\left( x\right) $ and cocycle $%
\Lambda _{ij}\left( x\right) \Lambda _{jk}\left( x\right) =\Lambda
_{ik}\left( x\right) $ consistency conditions. For trivial bundles, the
transition function reduces to%
\begin{equation}
\Lambda _{ij}\left( x\right) =g_{i}^{-1}g_{j}\text{,}  \label{transition}
\end{equation}%
where $g_{i}:\mathbb{F}\rightarrow \mathbb{F}$ is defined by $g_{i}:=\chi _{i%
\text{, }x}^{-1}\circ \widetilde{\chi }_{i\text{, }x}$ provided the local
trivializations $\left\{ \chi _{i}\right\} $ and $\left\{ \widetilde{\chi }%
_{i}\right\} $ give rise to the same fiber bundle.

A section is defined as a smooth map%
\begin{equation}
s:M\rightarrow \mathbb{P}\text{,}
\end{equation}%
such that $s(x)\in \pi ^{-1}\left( x\right) =\mathbb{F}_{x}$ $\forall x\in M$
and satisfies%
\begin{equation}
\pi \circ s=\left( id\right) _{M}\text{,}
\end{equation}%
where $\left( id\right) _{M}$ is the identity\ element of $M$. It assigns to
each point $x\in M$ a point in the fiber over $x$. Trivial bundles admit
global sections.

A bundle is a principal fiber bundle $\left\langle \mathbb{P}\text{, }%
\mathbb{P}/G\text{, }G\text{, }\pi \right\rangle $ provided the Lie group $G$
acts freely (i.e. if $pg=p$ then $g=\left( id\right) _{G}$) on $\mathbb{P}$
to the right $R_{g}p=pg$, $p\in \mathbb{P}$, preserves fibers on $\mathbb{P}$
($R_{g}:\mathbb{P}\rightarrow \mathbb{P}$), and is transitive on fibers.
Furthermore, there must exist local trivializations compatible with the $G$
action. Hence, $\pi ^{-1}(\mathcal{U}_{i})$ is homeomorphic to $\mathcal{U}%
_{i}\times _{M}G$ and the fibers of $\mathbb{P}$ are diffeomorphic to $G$.
The trivialization or inverse diffeomorphism map is given by%
\begin{equation}
\chi _{i}^{-1}:\pi ^{-1}(\mathcal{U}_{i})\rightarrow \mathcal{U}_{i}\times
_{M}G  \label{trivial}
\end{equation}%
such that $\chi ^{-1}(p)=\left( \pi (p)\text{, }\varphi (p)\right) \in
\mathcal{U}_{i}\times _{M}G$, $p\in \pi ^{-1}(\mathcal{U}_{i})\subset
\mathbb{P}$, where we see from the above definition that $\varphi $ is a
local mapping of $\pi ^{-1}(\mathcal{U}_{i})$ into $G$ satisfying $\varphi
(L_{g}p)$ $=\varphi (p)g$ for any $p\in \pi ^{-1}(\mathcal{U})$ and any $%
g\in G$. Observe that the elements of $\mathbb{P}$ which are projected onto
the same $x\in \left\{ \mathcal{U}_{i}\right\} $ are transformed into one
another by the elements of $G$. In other words, the fibers of $\mathbb{P}$
are the orbits of $G$ and at the same time, the set of elements which are
projected onto the same $x\in \mathcal{U}\subset M$. This observation
motivates calling the action of the group vertical\ and the base manifold
horizontal. The diffeomorphism map $\chi _{i}$ is called the local gauge
since $\chi _{i}^{-1}$ maps $\pi ^{-1}(\mathcal{U}_{i})$\ onto the direct
(Cartesian) product $\mathcal{U}_{i}\times _{M}G$. The action $L_{g}$ of the
structure group $G$ on $\mathbb{P}$ defines an isomorphism of the Lie
algebra $\mathfrak{g}$ of $G$ onto the Lie algebra of vertical vector fields
on $\mathbb{P}$ tangent to the fiber at each $p\in \mathbb{P}$ called
fundamental vector fields%
\begin{equation}
\lambda _{g}:T_{p}\left( \mathbb{P}\right) \rightarrow T_{gp}(\mathbb{P}%
)=T_{\pi (p)}\left( \mathbb{P}\right) \text{,}
\end{equation}%
where $T_{p}\left( \mathbb{P}\right) $ is the space of tangents at $p$, i.e.
$T_{p}\left( \mathbb{P}\right) \in T\left( \mathbb{P}\right) $. The map $%
\lambda $ is a linear isomorphism for every $p\in \mathbb{P}$ and is
invariant with respect to the action of $G$, that is, $\lambda _{g}:\left(
\lambda _{g\ast }T_{p}\left( \mathbb{P}\right) \right) \rightarrow
T_{gp}\left( \mathbb{P}\right) $, where $\lambda _{g\ast }$ is the
differential push forward map induced by $\lambda _{g}$ defined by $\lambda
_{g\ast }:T_{p}\left( \mathbb{P}\right) \rightarrow T_{gp}\left( \mathbb{P}%
\right) $.

Since the principal bundle $\mathbb{P}\left( M\text{, }G\right) $ is a
differentiable manifold, we can define tangent $T\left( \mathbb{P}\right) $
and cotangent $T^{\ast }\left( \mathbb{P}\right) $ bundles. The tangent
space $T_{p}\left( \mathbb{P}\right) $\ defined at each point $p\in \mathbb{P%
}$ may be decomposed into a vertical $V_{p}\left( \mathbb{P}\right) $ and
horizontal $H_{p}\left( \mathbb{P}\right) $ subspace as $T_{p}\left( \mathbb{%
P}\right) :=V_{p}\left( \mathbb{P}\right) \oplus H_{p}\left( \mathbb{P}%
\right) $ (where $\oplus $ represents the direct sum). The space $%
V_{p}\left( \mathbb{P}\right) $ is a subspace of $T_{p}\left( \mathbb{P}%
\right) $ consisting of all tangent vectors to the fiber passing through $%
p\in \mathbb{P}$, and $H_{p}\left( \mathbb{P}\right) $\ is the subspace
complementary to $V_{p}\left( \mathbb{P}\right) $ at $p$. The vertical
subspace $V_{p}\left( \mathbb{P}\right) :=\left\{ X\in T\left( \mathbb{P}%
\right) |\pi \left( X\right) \in \mathcal{U}_{i}\subset M\right\} $ is
uniquely determined by the structure of $\mathbb{P}$, whereas the horizontal
subspace $H_{p}\left( \mathbb{P}\right) $ cannot be uniquely specified. Thus
we require the following condition: when $p$ transforms as $p\rightarrow
p^{\prime }=pg$, $H_{p}\left( \mathbb{P}\right) $ transforms as \cite%
{Nakahara},%
\begin{equation}
R_{g\ast }H_{p}\left( \mathbb{P}\right) \rightarrow H_{p^{\prime }}\left(
\mathbb{P}\right) =R_{g}H_{p}\left( \mathbb{P}\right) =H_{pg}\left( \mathbb{P%
}\right) .
\end{equation}%
Let the local coordinates of $\mathbb{P}\left( M\text{, }G\right) $ be $%
p=\left( x\text{, }g\right) $ where $x\in M$ and $g\in G$. Let $\mathbf{G}%
_{A}$ denote the generators of the Lie algebra $\mathfrak{g}$ corresponding
to group $G$ satisfying the commutators $\left[ \mathbf{G}_{A}\text{, }%
\mathbf{G}_{B}\right] =f_{AB}^{\text{ \ \ \ }C}\mathbf{G}_{C}$, where $%
f_{AB}^{\text{ \ \ \ }C}$ are the structure constants of $G$. Let $\Omega $
be a connection form defined by $\Omega ^{A}:=\Omega _{i}^{A}dx^{i}\in
\mathfrak{g}$. Let $\omega $ be a connection 1-form defined by%
\begin{equation}
\omega :=\widetilde{g}^{-1}\pi _{\mathbb{P}M}^{\ast }\Omega \widetilde{g}+%
\widetilde{g}^{-1}d\widetilde{g}
\end{equation}%
($\ast $ represents the differential pullback map) belonging to $\mathfrak{g}%
\otimes T_{p}^{\ast }\left( \mathbb{P}\right) $ where $T_{p}^{\ast }\left(
\mathbb{P}\right) $ is the space dual to $T_{p}\left( \mathbb{P}\right) $.
The differential pullback map applied to a test function $\varphi $ and $p$%
-forms $\alpha $ and $\beta $ satisfy $f^{\ast }\varphi =\varphi \circ f$, $%
\left( g\circ f\right) ^{\ast }=f^{\ast }g^{\ast }$ and$\ f^{\ast }\left(
\alpha \wedge \beta \right) =f^{\ast }\alpha \wedge f^{\ast }\beta $. If $G$
is represented by a $d$-dimensional $d\times d$ matrix, then $\mathbf{G}_{A}=%
\left[ \mathbf{G}_{\alpha \beta }\right] $,\ $\widetilde{g}=\left[
\widetilde{g}^{\alpha \beta }\right] $, where $\alpha $, $\beta =1$, $2$, $3$%
,$...d$. Thus, $\omega $ assumes the form%
\begin{equation}
\omega _{\alpha }^{\text{ }\beta }=\left( \widetilde{g}^{-1}\right) _{\alpha
\gamma }d\widetilde{g}^{\gamma \beta }+\left( \widetilde{g}^{-1}\right)
_{\rho \gamma }\pi _{\mathbb{P}M}^{\ast }\Omega _{\text{ }\sigma i}^{\rho }%
\mathbf{G}_{\alpha }^{\text{ }\gamma }\widetilde{g}^{\sigma \beta }\otimes
dx^{i}\text{.}
\end{equation}

If $M$ is $n$-dimensional, the tangent space $T_{p}\left( \mathbb{P}\right) $
is $\left( n+d\right) $-dimensional. Since the vertical subspace $%
V_{p}\left( \mathbb{P}\right) $ is tangential to the fiber $G$, it is $d$%
-dimensional. Accordingly, $H_{p}\left( \mathbb{P}\right) $ is $n$%
-dimensional. The basis of $V_{p}\left( \mathbb{P}\right) $ can be taken to
be $\partial _{\alpha \beta }:=\frac{\partial }{\partial g^{\alpha \beta }}$%
. Now, let the basis of $H_{p}\left( \mathbb{P}\right) $ be denoted by%
\begin{equation}
E_{i}:=\partial _{i}+\Gamma _{i}^{\alpha \beta }\partial _{\alpha \beta }%
\text{,}\ i=1\text{, }2\text{, }3,..n\ \text{and}\ \alpha \text{, }\beta =1%
\text{, }2\text{, }3,..d
\end{equation}%
where $\partial _{i}=\frac{\partial }{\partial x^{i}}$. The connection
1-form $\omega $ projects $T_{p}\left( \mathbb{P}\right) $ onto $V_{p}\left(
\mathbb{P}\right) $. In order for $X\in T_{p}\left( \mathbb{P}\right) $ to
belong to $H_{p}\left( \mathbb{P}\right) $, that is for $X\in H_{p}\left(
\mathbb{P}\right) $, $\omega _{p}\left( X\right) =\left\langle \omega \left(
p\right) |X\right\rangle =0$. In other words,
\begin{equation}
H_{p}\left( \mathbb{P}\right) :=\left\{ X\in T_{p}\left( \mathbb{P}\right)
|\omega _{p}\left( X\right) =0\right\} \text{,}
\end{equation}%
from which $\Omega _{i}^{\alpha \beta }$ can be determined. The inner
product appearing in $\omega _{p}\left( X\right) =\left\langle \omega \left(
p\right) |X\right\rangle =0$ is a map $\left\langle \cdot |\cdot
\right\rangle :T_{p}^{\ast }\left( \mathbb{P}\right) \times T_{p}\left(
\mathbb{P}\right) \rightarrow
%TCIMACRO{\U{211d} }%
%BeginExpansion
\mathbb{R}
%EndExpansion
$ defined by $\left\langle W|V\right\rangle =W_{\mu }V^{\nu }\left\langle
dx^{\mu }|\frac{\partial }{\partial x^{\nu }}\right\rangle =W_{\mu }V^{\nu
}\delta _{\nu }^{\mu }$, where the 1-form $W$ and vector $V$ are given by $%
W=W_{\mu }dx^{\mu }$ and $V=V^{\mu }\frac{\partial }{\partial x^{\nu }}$.
Observe also that, $\left\langle dg^{\alpha \beta }|\partial _{\rho \sigma
}\right\rangle =\delta _{\rho }^{\alpha }\delta _{\sigma }^{\beta }$.

We parameterize an arbitrary group element $\widetilde{g}_{\lambda }$ as $%
\widetilde{g}\left( \lambda \right) =e^{\lambda ^{A}\mathbf{G}%
_{A}}=e^{\lambda \cdot \mathbf{G}}$,\ $A=1$,$..dim\left( \mathfrak{g}\right)
$. The right action $R_{\widetilde{g}\left( \lambda \right) }=R_{\exp \left(
\lambda \cdot G\right) }$ on $p\in \mathbb{P}$, i.e. $R_{\exp \left( \lambda
\cdot \mathbf{G}\right) }p=p\exp \left( \lambda \cdot \mathbf{G}\right) $,
defines a curve through $p$ in $\mathbb{P}$. Define a vector $G^{\#}\in
T_{p}\left( \mathbb{P}\right) $ by \cite{Nakahara}%
\begin{equation}
G^{\#}f\left( p\right) :=\frac{d}{dt}f\left( p\exp \left( \lambda \cdot
\mathbf{G}\right) \right) |_{\lambda =0}
\end{equation}%
where $f:\mathbb{P}\rightarrow
%TCIMACRO{\U{211d} }%
%BeginExpansion
\mathbb{R}
%EndExpansion
$ is an arbitrary smooth function. Since the vector $G^{\#}$ is tangent to $%
\mathbb{P}$ at $p$, $G^{\#}\in V_{p}\left( \mathbb{P}\right) $, the
components of the vector $G^{\#}$ are the fundamental vector fields at $p$
which constitute $V(\mathbb{P})$. The components of $G^{\#}$ may also be
viewed as a basis element of the Lie algebra $\mathfrak{g}$. Given $%
G^{\#}\in V_{p}\left( \mathbb{P}\right) $, $\mathbf{G}\in \mathfrak{g}$,%
\begin{eqnarray}
\omega _{p}\left( G^{\#}\right) &=&\left\langle \omega \left( p\right)
|G^{\#}\right\rangle =\widetilde{g}^{-1}d\widetilde{g}\left( G^{\#}\right) +%
\widetilde{g}^{-1}\pi _{\mathbb{P}M}^{\ast }\Omega \widetilde{g}\left(
G^{\#}\right)  \notag \\
&=&\widetilde{g}_{p}^{-1}\widetilde{g}_{p}\frac{d}{d\lambda }\left( \exp
\left( \lambda \cdot \mathbf{G}\right) \right) |_{\lambda =0}\text{,}
\end{eqnarray}%
where use was made of $\pi _{\mathbb{P}M\ast }G^{\#}=0$. Hence, $\omega
_{p}\left( G^{\#}\right) =\mathbf{G}$. An arbitrary vector $X\in H_{p}\left(
\mathbb{P}\right) $ may be expanded in a basis spanning $H_{p}\left( \mathbb{%
P}\right) $ as $X:=\beta ^{i}E_{i}$. By direct computation, one can show%
\begin{equation}
\left\langle \omega _{\alpha }^{\text{ }\beta }|X\right\rangle =\left(
\widetilde{g}^{-1}\right) _{\alpha \gamma }\beta ^{i}\Gamma _{i}^{\gamma
\beta }+\left( \widetilde{g}^{-1}\right) _{\alpha \gamma }\pi _{\mathbb{P}%
M}^{\ast }\Omega _{\text{ }\sigma i}^{\rho }\beta ^{i}\mathbf{G}_{\rho
}^{\gamma }\widetilde{g}^{\sigma \beta }=0\text{, }\forall \beta ^{i}
\label{conn-vect}
\end{equation}%
Equation (\ref{conn-vect}) yields%
\begin{equation}
\left( \widetilde{g}^{-1}\right) _{\alpha \gamma }\Gamma _{i}^{\gamma \beta
}+\left( \widetilde{g}^{-1}\right) _{\alpha \gamma }\pi _{\mathbb{P}M}^{\ast
}\Omega _{\text{ }\sigma i}^{\rho }\mathbf{G}_{\rho }^{\gamma }\widetilde{g}%
^{\sigma \beta }=0\text{,}
\end{equation}%
from which we obtain%
\begin{equation}
\Gamma _{i}^{\gamma \beta }=-\pi _{\mathbb{P}M}^{\ast }\Omega _{\text{ }%
\sigma i}^{\rho }\mathbf{G}_{\rho }^{\gamma }\widetilde{g}^{\sigma \beta }%
\text{.}
\end{equation}%
In this manner, the horizontal component is completely determined. An
arbitrary tangent vector $\mathfrak{X}\in T_{p}\left( \mathbb{P}\right) $
defined at $p\in \mathbb{P}$ takes the form%
\begin{equation}
\mathfrak{X}=A^{\alpha \beta }\partial _{\alpha \beta }+B^{i}\left( \partial
_{i}-\pi _{\mathbb{P}M}^{\ast }\Omega _{\text{ }\sigma i}^{\rho }\mathbf{G}%
_{\rho }^{\alpha }\widetilde{g}^{\sigma \beta }\partial _{\alpha \beta
}\right) ,
\end{equation}%
where $A^{\alpha \beta }$ and $B^{i}$ are constants. The vector field $%
\mathfrak{X}$ is comprised of horizontal $\mathfrak{X}^{H}:=B^{i}\left(
\partial _{i}-\pi _{\mathbb{P}M}^{\ast }\Omega _{\text{ }\sigma i}^{\rho }%
\mathbf{G}_{\rho }^{\alpha }\widetilde{g}^{\sigma \beta }\partial _{\alpha
\beta }\right) \in H\left( \mathbb{P}\right) $ and vertical $\mathfrak{X}%
^{V}:=A^{\alpha \beta }\partial _{\alpha \beta }\in V\left( \mathbb{P}%
\right) $ components.

Let $\mathfrak{X}\in T_{p}\left( \mathbb{P}\right) $ and $g\in \mathbf{G}$,\
then%
\begin{equation}
R_{g}^{\ast }\omega \left( \mathfrak{X}\right) =\omega \left( R_{g\ast }%
\mathfrak{X}\right) =\widetilde{g}_{pg}^{-1}\Omega \left( R_{g\ast }%
\mathfrak{X}\right) \widetilde{g}_{pg}+\widetilde{g}_{pg}^{-1}d\widetilde{g}%
_{pg}\left( R_{g\ast }\mathfrak{X}\right) \text{,}  \label{Rightw}
\end{equation}%
Observing that $\widetilde{g}_{pg}=\widetilde{g}_{p}g$ and $\widetilde{g}%
_{gp}^{-1}=g^{-1}\widetilde{g}_{p}^{-1}$ the first term on the RHS of (\ref%
{Rightw}) reduces to $\widetilde{g}_{pg}^{-1}\Omega \left( R_{g\ast }%
\mathfrak{X}\right) \widetilde{g}_{pg}=g^{-1}\widetilde{g}_{p}^{-1}\Omega
\left( R_{g\ast }\mathfrak{X}\right) \widetilde{g}_{p}g$ while the second
term gives $\widetilde{g}_{pg}^{-1}d\widetilde{g}_{pg}\left( R_{g\ast }%
\mathfrak{X}\right) =g^{-1}\widetilde{g}_{p}^{-1}d\left( R_{g\ast }\mathfrak{%
X}\right) \widetilde{g}_{p}g$. We therefore conclude%
\begin{equation}
R_{g}^{\ast }\omega _{\lambda }=ad_{g^{-1}}\omega _{\lambda }\text{,}
\end{equation}%
where the adjoint map $ad$ is defined by%
\begin{equation}
ad_{g}Y:=L_{g\ast }\circ R_{g^{-1}\ast }\circ Y=gYg^{-1}\text{, \ }%
ad_{g^{-1}}Y:=g^{-1}Yg\text{.}
\end{equation}

The potential $\Omega ^{A}$ can be obtained from $\omega $ as $\Omega
^{A}=s^{\ast }\omega $. To demonstrate this, let $Y\in T_{p}\left( M\right) $
and $\widetilde{g}$ be specified by the inverse diffeomorphism or
trivialization map (\ref{trivial}) with $\chi _{\lambda }^{-1}\left(
p\right) =\left( x\text{, }\widetilde{g}_{\lambda }\right) $ for $p\left(
x\right) =s_{\lambda }\left( x\right) \cdot \widetilde{g}_{\lambda }$. We
find $s_{i}^{\ast }\omega \left( Y\right) =\widetilde{g}^{-1}\Omega \left(
\pi _{\ast }s_{i\ast }Y\right) \widetilde{g}+\widetilde{g}^{-1}d\widetilde{g}%
\left( s_{i\ast }Y\right) $, where we \cite{Nakahara} have used $s_{i\ast
}Y\in T_{s_{i}}\left( \mathbb{P}\right) $, $\pi _{\ast }s_{i\ast }=\left(
id\right) _{T_{p}\left( M\right) }$ and $\widetilde{g}=\left( id\right) _{G}$
at $s_{i}$ implying $\widetilde{g}^{-1}d\widetilde{g}\left( s_{i\ast
}Y\right) =0$. Hence,%
\begin{equation}
s_{i}^{\ast }\omega \left( Y\right) =\Omega \left( Y\right) \text{.}
\end{equation}

To determine the gauge transformation of the connection 1-form $\omega $ we
use the fact that $R_{\widetilde{g}\ast }X=X\widetilde{g}$ for $X\in
T_{p}\left( M\right) $ and the transition functions $\widetilde{g}_{nm}\in G$
defined between neighboring bundle charts (\ref{transition}). By direct
computation we get%
\begin{eqnarray}
c_{j\ast }X &=&\frac{d}{dt}c_{j}\left( \lambda \left( t\right) \right)
|_{t=0}=\frac{d}{dt}\left[ c_{i}\left( \lambda \left( t\right) \right) \cdot
\widetilde{g}_{ij}\right] |_{t=0}  \notag \\
&=&R_{\widetilde{g}_{ij}\ast }c_{i}^{\ast }\left( X\right) +\left(
\widetilde{g}_{ji}^{-1}\left( x\right) d\widetilde{g}_{ij}\left( X\right)
\right) ^{\#}\text{.}
\end{eqnarray}%
where $\lambda \left( t\right) $ is a curve in $M$ with boundary values $%
\lambda \left( 0\right) =m$ and $\frac{d}{dt}\lambda \left( t\right)
|_{t=0}=X$. Thus, we obtain the useful result%
\begin{equation}
c_{\ast }X=R_{\widetilde{g}\ast }\left( c_{\ast }X\right) +\left( \widetilde{%
g}^{-1}d\widetilde{g}\left( X\right) \right) ^{\#}\text{.}  \label{trans1}
\end{equation}%
Applying $\omega $ to (\ref{trans1}) we get%
\begin{equation}
\omega \left( c_{\ast }X\right) =c^{\ast }\omega \left( X\right) =ad_{%
\widetilde{g}^{-1}}c^{\ast }\omega \left( X\right) +\widetilde{g}^{-1}d%
\widetilde{g}\left( X\right) \text{, }\forall X\text{.}
\end{equation}%
Hence, the gauge transformation of the local gauge potential $\Omega $ reads,%
\begin{equation}
\Omega \rightarrow \Omega ^{\prime }=ad_{\widetilde{g}^{-1}}\left( d+\Omega
\right) =\widetilde{g}^{-1}\left( d+\Omega \right) \widetilde{g}\text{.}
\label{Lconn-trans}
\end{equation}%
Since $\Omega =c^{\ast }\omega $ we obtain from (\ref{Lconn-trans}) the
gauge transformation law of $\omega $%
\begin{equation}
\omega \rightarrow \omega ^{\prime }=\widetilde{g}^{-1}\left( d+\omega
\right) \widetilde{g}\text{.}
\end{equation}

\section{Generalized Bundle Structure of Gravitation}

Let us recall the definition of gauge transformations in the context of
ordinary fiber bundles. Given a principal fiber bundle $\mathbb{P}(M$, $G$; $%
\pi )$ with base space $M$ and standard $G$-diffeomorphic fiber, gauge
transformations are characterized by bundle isomorphisms \cite{Giachetta} $%
\lambda :\mathbb{P}\rightarrow \mathbb{P}$ exhausting all diffeomorphisms $%
\lambda _{M}$ on $M$. This mapping is called an automorphism of $\mathbb{P}$
provided it is equivariant with respect to the action of $G$. This amounts
to restricting the action $\lambda $ of $G$ along local fibers leaving the
base space unaffected. Indeed, with regard to gauge theories of internal
symmetry groups, a gauge transformation is a fiber preserving bundle
automorphism, i.e. diffeomorphisms $\lambda $ with $\lambda _{M}=\left(
id\right) _{M}$. The automorphisms $\lambda $ form a group called the
automorphism group $Aut_{\mathbb{P}}$ of $\mathbb{P}$. The gauge
transformations form a subgroup of $Aut_{\mathbb{P}}$ called the gauge group
$G\left( Aut_{\mathbb{P}}\right) $ (or $G$ in short) of $\mathbb{P}$.

The map $\lambda $ is required to satisfy two conditions, namely its
commutability with the right action of $G$ $[$the equivariance condition $%
\lambda \left( R_{g}(p)\right) =\lambda \left( pg\right) =\lambda \left(
p\right) g]$%
\begin{equation}
\lambda \circ R_{g}(p)=R_{g}(p)\circ \lambda \text{, \ }p\in \mathbb{P}\text{%
, }g\in G
\end{equation}%
according to which fibers are mapped into fibers, and the verticality
condition%
\begin{equation}
\pi \circ \lambda \left( u\right) =\pi \left( u\right) \text{,}
\end{equation}%
where $u$ and $\lambda \left( u\right) $ belong to the same fiber. The last
condition ensures that no diffeomorphisms $\lambda _{M}:M\rightarrow M$
given by%
\begin{equation}
\lambda _{M}\circ \pi \left( u\right) =\pi \circ \lambda \left( u\right)
\text{,}
\end{equation}%
be allowed on the base space $M$. In a gauge description of gravitation, one
is interested in gauging external transformation groups. That is to say the
group action on spacetime coordinates cannot be neglected. The spaces of
internal fiber and external base must be interlocked in the sense that
transformations in one space must induce corresponding transformations in
the other. The usual definition of a gauge transformation, i.e. as a
displacement along local fibers not affecting the base space, must be
generalized to reflect this interlocking. One possible way of framing this
interlocking is to employ a nonlinear realization of the gauge group $G$,
provided a closed subgroup $H\subset G$ exist. The interlocking requirement
is then transformed into the interplay between groups $G$ and one of its
closed subgroups $H$.

Denote by $G$ a Lie group with elements $\left\{ g\right\} $. Let $H$ be a
closed subgroup of $G$ specified by $[37$, $67]$%
\begin{equation}
H:=\left\{ h\in G|\Pi \left( R_{h}g\right) =\pi \left( g\right) \text{, }%
\forall g\in G\right\} \text{,}
\end{equation}%
with elements $\left\{ h\right\} $ and known linear representations $\rho
\left( h\right) $. Here $\Pi $ is the first of the two projection maps in (%
\ref{comp-pro}), and $R_{h}$ is the right group action. Let $M$ be a
differentiable manifold with points $\left\{ x\right\} $ to which $G$ and $H$
may be referred, i.e. $g=g(x)$ and $h=h(x)$. Being that $G$ and $H$ are Lie
groups, they are also manifolds. The right action of $H$ on $G$ induce a
complete partition of $G$ into mutually disjoint orbits $gH$. Since $g=g(x)$%
, all elements of $gH=\left\{ gh_{1}\text{, }gh_{2}\text{, }gh_{3}\text{,}%
\cdot \cdot \cdot \text{, }gh_{n}\right\} $ are defined over the same $x$.
Thus, each orbit $gH$ constitute an equivalence class of point $x$, with
equivalence relation $g\equiv g^{\prime }$ where\ $g^{\prime }=R_{h}g=gh$.
By projecting each equivalence class onto a single element of the quotient
space $\mathcal{M}:=G/H$, the group $G$ becomes organized as a fiber bundle
\ in\ the sense that $G=\bigcup\nolimits_{i}\left\{ g_{i}H\right\} $. In
this manner the manifold $G$ is viewed as a fiber bundle $G\left( \mathcal{M}%
\text{, }H\text{; }\Pi \right) $ with $H$-diffeomorphic fibers $\Pi
^{-1}\left( \xi \right) :G\rightarrow \mathcal{M}=gH$ and base space $%
\mathcal{M}$. A composite principal fiber bundle\textit{\ }$\mathbb{P}(M$, $G
$; $\pi )$ is one whose $G$-diffeomorphic fibers possess the fibered
structure $G\left( \mathcal{M}\text{, }H\text{; }\Pi \right) \simeq \mathcal{%
M}\times $ $H$ described above. The bundle $\mathbb{P}$ is then locally
isomorphic to $M\times G\left( \mathcal{M}\text{, }H\right) $. Moreover,
since an element $g\in G$ is locally homeomorphic to $\mathcal{M}\times H$
the elements of $\mathbb{P}$ are - by transitivity - also locally
homeomorphic to $M\times \mathcal{M}\times H\simeq \Sigma \times H$ where
(locally) $\Sigma \simeq M\times \mathcal{M}$. Thus, an alternative view
\cite{Tresguerres} of $\mathbb{P}(M$, $G$; $\pi )$ is provided by the $%
\mathbb{P}$-associated $H$-bundle $\mathbb{P}(\Sigma $, $H$; $\widetilde{\pi
})$. The total space $\mathbb{P}$ may be regarded as $G\left( \mathcal{M}%
\text{, }H\text{; }\Pi \right) $-bundles over base space $M$ or equivalently
as $H$-fibers attached to manifold $\Sigma \simeq M\times \mathcal{M}$.

The nonlinear realization (NLR) technique \cite{CCWZ1, CCWZ2} provides a way
to determine the transformation properties of fields defined on the quotient
space $G/H$. The NLR of Diff$\left( 4\text{, }%
%TCIMACRO{\U{211d} }%
%BeginExpansion
\mathbb{R}
%EndExpansion
\right) $ becomes tractable due to a theorem given by V. I. Ogievetsky.
According to the Ogievetsky theorem \cite{BorisovOgievetskii}, the algebra
of the infinite dimensional group Diff$\left( 4\text{, }%
%TCIMACRO{\U{211d} }%
%BeginExpansion
\mathbb{R}
%EndExpansion
\right) $ can be taken as the closure of the finite dimensional algebras of $%
SO(4$, $2)$ and $A(4$, $%
%TCIMACRO{\U{211d} }%
%BeginExpansion
\mathbb{R}
%EndExpansion
)$. Remind that the Lorentz group generates transformations that preserve
the quadratic form on Minkowski spacetime built from the metric tensor,
while the special conformal group generates infinitesimal angle-preserving
transformations on Minkowski spacetime. The affine group is a generalization
of the Poincar\'{e} group where the Lorentz group is replaced by the group
of general linear transformations. As such, the affine group generates
translations, Lorentz transformations, volume preserving shear and volume
changing dilation transformations. As a consequence, the NLR of Diff$\left( 4%
\text{, }%
%TCIMACRO{\U{211d} }%
%BeginExpansion
\mathbb{R}
%EndExpansion
\right) /SO(3$, $1)$ can be constructed by taking a simultaneous realization
of the conformal group $SO(4$, $2)$ and the affine group $A(4$, $%
%TCIMACRO{\U{211d} }%
%BeginExpansion
\mathbb{R}
%EndExpansion
):=%
%TCIMACRO{\U{211d} }%
%BeginExpansion
\mathbb{R}
%EndExpansion
^{4}\rtimes GL(4$, $%
%TCIMACRO{\U{211d} }%
%BeginExpansion
\mathbb{R}
%EndExpansion
)$ on the coset spaces $A(4$, $%
%TCIMACRO{\U{211d} }%
%BeginExpansion
\mathbb{R}
%EndExpansion
)/SO(3$, $1)$ and $SO(4$, $2)/SO(3$, $1)$. One possible interpretation of
this theorem is that the conform-affine group (defined below) may be the
largest subgroup of Diff$\left( 4\text{, }%
%TCIMACRO{\U{211d} }%
%BeginExpansion
\mathbb{R}
%EndExpansion
\right) $ whose transformations may be put into the form of a generalized
coordinate transformation. We remark that a NLR can be made linear by
embedding the representation in a sufficiently higher dimensional space.
Alternatively, a linear group realization becomes nonlinear when subject to
constraints. One type of relevant constraints may be those responsible for
symmetry reduction from Diff$\left( 4\text{, }%
%TCIMACRO{\U{211d} }%
%BeginExpansion
\mathbb{R}
%EndExpansion
\right) $ to $SO(3$, $1)$ for instance.

We take the group $CA(3$, $1)$ as the basic symmetry group $G$. The CA group
consists of the groups $SO(4$, $2)$ and $A(4$, $%
%TCIMACRO{\U{211d} }%
%BeginExpansion
\mathbb{R}
%EndExpansion
)$. In particular, CA is proportional to the union $SO(4$, $2)\cup A(4$, $%
%TCIMACRO{\U{211d} }%
%BeginExpansion
\mathbb{R}
%EndExpansion
)$. We know however (see section \textit{Conform-Affine Lie Algebra}) that
the affine and special conformal groups have several group generators in
common. These common generators reside in the intersection $SO(4$, $2)\cap
A(4$, $%
%TCIMACRO{\U{211d} }%
%BeginExpansion
\mathbb{R}
%EndExpansion
)$ of the two groups, within which there are \textit{two copies} of $\Pi
:=D\times P(3$, $1)$, where $D$ is the group of scale transformations
(dilations)\ and $P(3$, $1):=T\left( 3\text{, }1\right) \rtimes SO(3$, $1)$
is the Poincar\'{e} group. We define the CA group as the union of the affine
and conformal groups minus \textit{one copy} of the overlap $\Pi $, i.e. $%
CA(3$, $1):=SO(4$, $2)\cup A(4$, $%
%TCIMACRO{\U{211d} }%
%BeginExpansion
\mathbb{R}
%EndExpansion
)-\Pi $. Being defined in this way we recognize that $CA(3$, $1)$ is a $24$
parameter Lie group representing the action of Lorentz transformations $(6)$%
, translations $(4)$, special conformal transformations $(4)$, spacetime
shears $(9)$ and scale transformations $(1)$. In this paper, we obtain the
NLR of $CA(3$, $1)$ modulo $SO(3$, $1)$.

\subsection{Conform-Affine Lie Algebra}

In order to implement the NLR\ procedure, we choose to partition Diff$(4$, $%
%TCIMACRO{\U{211d} }%
%BeginExpansion
\mathbb{R}
%EndExpansion
)$ with respect to the Lorentz group. By Ogievetsky's theorem \cite%
{BorisovOgievetskii}, we identify representations of Diff$(4$, $%
%TCIMACRO{\U{211d} }%
%BeginExpansion
\mathbb{R}
%EndExpansion
)/SO(3$, $1)$ with those of $CA(3$, $1)/SO(3$, $1)$. The $20$ generators of
affine transformations can be decomposed into the $4$ translational $\mathbf{%
P}_{\mu }^{\text{Aff}}$ and $16$ $GL(4$, $%
%TCIMACRO{\U{211d} }%
%BeginExpansion
\mathbb{R}
%EndExpansion
)$ transformations $\mathbf{\Lambda }_{\alpha }^{\text{ }\beta }$. The $16$
generators $\mathbf{\Lambda }_{\alpha }^{\text{ }\beta }$ may be further
decomposed into the $6$ Lorentz generators $\mathbf{L}_{\alpha }^{\text{ }%
\beta }$ plus the remaining $10$ generators of symmetric linear
transformation $\mathbf{S}_{\alpha }^{\text{ }\beta }$, that is, $\mathbf{%
\Lambda }_{\text{ }\beta }^{\alpha }=\mathbf{L}_{\text{ }\beta }^{\alpha }+%
\mathbf{S}_{\text{ }\beta }^{\alpha }$. The $10$ parameter symmetric linear
generators $\mathbf{S}_{\alpha }^{\text{ }\beta }$ can be factored into the $%
9$ parameter shear (the traceless part of $\mathbf{S}_{\alpha }^{\text{ }%
\beta }$) generator defined by $^{\dagger }\mathbf{S}_{\alpha }^{\text{ }%
\beta }=\mathbf{S}_{\alpha }^{\text{ }\beta }-\frac{1}{4}\delta _{\alpha }^{%
\text{ }\beta }\mathbf{D}$, and the $1$ parameter dilaton generator $\mathbf{%
D}=tr\left( \mathbf{S}_{\alpha }^{\text{ }\beta }\right) $. Shear
transformations generated by $^{\dagger }\mathbf{S}_{\alpha }^{\text{ }\beta
}$ describe shape changing, volume preserving deformations, while the
dilaton generator gives rise to volume changing transformations. The four
diagonal elements of $\mathbf{S}_{\alpha }^{\text{ }\beta }$ correspond to
the generators of projective transformations. The $15$ generators of
conformal transformations are defined in terms of the set $\left\{
J_{AB}\right\} $ where $A=0$, $1$, $2$,..$5$. The elements $J_{AB}$ can be
decomposed into translations $\mathbf{P}_{\mu }^{\text{Conf}}:=J_{5\mu
}+J_{6\mu }$, special conformal generators $\mathbf{\Delta }_{\mu }:=J_{5\mu
}-J_{6\mu }$, dilatons $\mathbf{D}:=J_{56}$ and the Lorentz generators $%
\mathbf{L}_{\alpha \beta }:=J_{\alpha \beta }$. The Lie algebra of $CA(3$, $%
1)$ is characterized by the commutation relations%
\begin{equation}
\begin{array}{c}
\left[ \mathbf{\Lambda }_{\alpha \beta }\text{, }\mathbf{D}\right] =\left[
\mathbf{\Delta }_{\alpha }\text{, }\mathbf{\Delta }_{\beta }\right] =0\text{%
, }\left[ \mathbf{P}_{\alpha }\text{, }\mathbf{P}_{\beta }\right] =\left[
\mathbf{D}\text{, }\mathbf{D}\right] =0\text{,} \\
\left[ \mathbf{L}_{\alpha \beta }\text{, }\mathbf{P}_{\mu }\right] =io_{\mu
\lbrack \alpha }\mathbf{P}_{\beta ]}\text{, }\left[ \mathbf{L}_{\alpha \beta
}\text{, }\mathbf{\Delta }_{\gamma }\right] =io_{[\alpha |\gamma }\mathbf{%
\Delta }_{|\beta ]}\text{,} \\
\left[ \mathbf{\Lambda }_{\text{ }\beta }^{\alpha }\text{, }\mathbf{P}_{\mu }%
\right] =i\delta _{\mu }^{\alpha }\mathbf{P}_{\beta }\text{, }\left[ \mathbf{%
\Lambda }_{\text{ }\beta }^{\alpha }\text{, }\mathbf{\Delta }_{\mu }\right]
=i\delta _{\mu }^{\alpha }\mathbf{\Delta }_{\beta }\text{,} \\
\left[ \mathbf{S}_{\alpha \beta }\text{, }\mathbf{P}_{\mu }\right] =io_{\mu
(\alpha }\mathbf{P}_{\beta )}\text{, }\left[ \mathbf{P}_{\alpha }\text{, }%
\mathbf{D}\right] =-i\mathbf{P}_{\alpha }\text{,} \\
\left[ \mathbf{L}_{\alpha \beta }\text{, }\mathbf{L}_{\mu \nu }\right]
=-i\left( o_{\alpha \lbrack \mu }\mathbf{L}_{\nu ]\beta }-o_{\beta \lbrack
\mu }\mathbf{L}_{\nu ]\alpha }\right) \text{,} \\
\left[ \mathbf{S}_{\alpha \beta }\text{, }\mathbf{S}_{\mu \nu }\right]
=i\left( o_{\alpha (\mu }\mathbf{L}_{\nu )\beta }-o_{\beta (\mu }\mathbf{L}%
_{\nu )\alpha }\right) \text{,} \\
\left[ \mathbf{L}_{\alpha \beta }\text{, }\mathbf{S}_{\mu \nu }\right]
=i\left( o_{\alpha (\mu }\mathbf{S}_{\nu )\beta }-o_{\beta (\mu }\mathbf{S}%
_{\nu )\alpha }\right) \text{,} \\
\left[ \mathbf{\Delta }_{\alpha }\text{, }\mathbf{D}\right] =i\mathbf{\Delta
}_{\alpha }\text{, }\left[ \mathbf{S}_{\mu \nu }\text{, }\mathbf{\Delta }%
_{\alpha }\right] =io_{\alpha (\mu }\mathbf{\Delta }_{\nu )}\text{,} \\
\left[ \mathbf{\Lambda }_{\text{ }\beta }^{\alpha }\text{, }\mathbf{\Lambda }%
_{\text{ }\nu }^{\mu }\right] =i\left( \delta _{\nu }^{\alpha }\mathbf{%
\Lambda }_{\text{ }\beta }^{\mu }-\delta _{\beta }^{\mu }\mathbf{\Lambda }_{%
\text{ }\nu }^{\alpha }\right) \text{,} \\
\left[ \mathbf{P}_{\alpha }\text{, }\mathbf{\Delta }_{\beta }\right]
=2i\left( o_{\alpha \beta }\mathbf{D}-\mathbf{L}_{\alpha \beta }\right)
\text{, }%
\end{array}%
\end{equation}%
where $o_{\alpha \beta }=diag\left( -1\text{, }1\text{, }1\text{, }1\right) $
is Lorentz group metric.

\section{Group Actions and Bundle Morphisms}

In this section we introduce the main ingredients required to specify the
structure of the fiber bundle we employ, namely the canonical projection,
sections etc. Our main guide in this section is Tresguerres \cite%
{Tresguerres}. We follow his prescription for constructing the composite
fiber bundle, but implement the program for the CA group.

The composite bundle\textit{\ }$\mathbb{P}(\Sigma $, $H$; $\widetilde{\pi })$
is comprised of $H$-fibers, base space $\Sigma \left( M\text{, }\mathcal{M}%
\right) $ and a composite map
\begin{equation}
\widetilde{\pi }\overset{\text{def}}{=}\widetilde{\pi }_{\Sigma M}\circ \Pi
_{\mathbb{P}\Sigma }:\mathbb{P}\rightarrow \Sigma \rightarrow M\text{,}
\end{equation}%
with component projections%
\begin{equation}
\Pi _{\mathbb{P}\Sigma }:\mathbb{P}\rightarrow \Sigma \text{, \ }\widetilde{%
\pi }_{\Sigma M}:\Sigma \rightarrow M\text{.}  \label{comp-pro}
\end{equation}%
The projection $\Pi _{\mathbb{P}\Sigma }$ maps the point $\left( p\in
\mathbb{P}\text{, }R_{h}p\in \mathbb{P}\right) $ into point $\left( x\text{,
}\xi \right) \in $ $\Sigma $. There is a correspondence between sections $%
s_{M\Sigma }:M\rightarrow \Sigma $ and the projection $\Pi _{\mathbb{P}%
\Sigma }:\mathbb{P}\rightarrow \Sigma $ in the sense that both maps project
their functional argument onto elements of $\Sigma $. This is formalized by
the relation, $\Pi _{\mathbb{P}\Sigma }\left( p\right) =s_{M\Sigma }\circ
\pi _{\mathbb{P}M}\left( p\right) $. Hence, the total projection is given by%
\begin{equation}
\widetilde{\pi }:=\pi _{\mathbb{P}M}=\widetilde{\pi }_{\Sigma M}\circ \Pi _{%
\mathbb{P}\Sigma }.
\end{equation}%
Associated with the projections $\widetilde{\pi }_{\Sigma M}$ and $\Pi _{%
\mathbb{P}\Sigma }$ are the corresponding local sections%
\begin{equation}
s_{M\Sigma }:\mathcal{U}\rightarrow \widetilde{\pi }_{\Sigma M}^{-1}\left(
\mathcal{U}\right) \subset \Sigma \text{, }s_{\Sigma \mathbb{P}}:\mathcal{V}%
\rightarrow \Pi _{\mathbb{P}\Sigma }^{-1}\left( \mathcal{V}\right) \subset
\mathbb{P}\text{,}
\end{equation}%
with neighborhoods $\mathcal{U}\subset M$ and $\mathcal{V}\subset \Sigma $
satisfying%
\begin{equation}
\widetilde{\pi }_{\Sigma M}\circ s_{M\Sigma }=\left( id\right) _{M}\text{, }%
\Pi _{\mathbb{P}\Sigma }\circ s_{\Sigma \mathbb{P}}=\left( id\right)
_{\Sigma }\text{.}
\end{equation}%
The bundle injection $\widetilde{\pi }^{-1}\left( \mathcal{U}\right) $ is
the inverse image of $\widetilde{\pi }\left( \mathcal{U}\right) $ and is
called the fiber over $\mathcal{U}$. The equivalence class $R_{h}p=pH\in
\widetilde{\pi }_{\Sigma M}^{-1}\left( \mathcal{U}\right) $ of left cosets
is the fiber of $\mathbb{P}\left( \Sigma \text{, }H\right) $ while each
orbit $pH$ through $p\in \mathbb{P}$\ projects into a single element $Q\in $
$\Sigma $. In analogy to the total bundle projection (\ref{comp-pro}), a
total section of $\mathbb{P}$ is given by the total section composition%
\begin{equation}
s_{M\mathbb{P}}=s_{\Sigma \mathbb{P}}\circ s_{M\Sigma }.  \label{comp-sect}
\end{equation}%
Let elements of $G/H$ be labeled by the parameter $\xi $. Functions on $G/H$
are represented by continuous coset functions $c(\xi )$ parameterized by $%
\xi $. These elements are referred to as cosets to the right of $H$ with
respect to $g\in G$. Indeed, the orbits of the right action of $H$ on $G$
are the left cosets $R_{h}g=gH$. For a given section $s_{M\mathbb{P}}\left(
x\in M\right) \in \pi _{\mathbb{P}M}^{-1}$ with local coordinates $\left( x%
\text{, }g\right) $ one can perform decompositions of the partial fibers $%
s_{M\Sigma }$ and $s_{\Sigma \mathbb{P}}$ as:
\begin{equation}
s_{M\Sigma }\left( x\right) =\widetilde{c}_{M\Sigma }\left( x\right) \cdot
c=R_{c^{\prime }}\circ \widetilde{c}_{M\Sigma }\left( x\right) \text{; }%
c=c\left( \xi \right) \text{,}  \label{sect-Msig}
\end{equation}%
\begin{equation}
s_{\Sigma \mathbb{P}}\left( x\text{, }\xi \right) =\widetilde{c}_{\Sigma
\mathbb{P}}\left( x\text{, }\xi \right) \cdot a^{\prime }=R_{a^{\prime
}}\circ \widetilde{c}_{\Sigma \mathbb{P}}\left( x\text{, }\xi \right) \text{%
; }a^{\prime }\in H\text{,}  \label{sect-sigP}
\end{equation}%
with the null sections $\left\{ \widetilde{c}_{M\Sigma }\left( x\right)
\right\} $ and $\left\{ \widetilde{c}_{\Sigma \mathbb{P}}\left( x\text{, }%
\xi \right) \right\} $ having coordinates $\left( x\text{, }\left( id\right)
_{\mathcal{M}}\right) $ and $\left( x\text{, }\xi \text{, }\left( id\right)
_{H}\right) $ respectively. A null or zero section is a map that sends every
point $x\in M$ to the origin of the fiber $\pi ^{-1}\left( x\right) $ over $%
x $, i.e. $\chi _{i}^{-1}\left( \widetilde{c}\left( x\right) \right) =\left(
x\text{, }0\right) $ in any trivialization. The trivialization map $\chi
_{i}^{-1}$ is defined in (\ref{trivial}) The identity map appearing in the
above trivializations are defined as $\left( id\right) _{\mathcal{M}}:%
\mathcal{M}\rightarrow \mathcal{M}$ and\ $\left( id\right) _{H}:H\rightarrow
H$. We assume the total null bundle section be given by the composition law
\begin{equation}
\widetilde{c}_{M\mathbb{P}}=\widetilde{c}_{\Sigma \mathbb{P}}\circ
\widetilde{c}_{M\Sigma }\text{.}  \label{comp-nullsect}
\end{equation}%
The images of two sections $s_{\Sigma \mathbb{P}}$ and $s_{M\Sigma }$ over $%
x\in M$ must coincide, implying $s_{\Sigma \mathbb{P}}\left(
x\text{, }\xi \right) =s_{M\Sigma }\left( x\right) $. Using
(\ref{comp-sect}) with (\ref{sect-Msig}), (\ref{sect-sigP}) and
(\ref{comp-nullsect}), we arrive at the total bundle section
decomposition
\begin{equation}
s_{M\mathbb{P}}\left( x\right) =\widetilde{c}_{M\mathbb{P}}\left( x\right)
\cdot g=R_{g}\circ \widetilde{c}_{M\mathbb{P}}\left( x\right)
\label{sMP-null}
\end{equation}%
provided $g=c\cdot a$ and
\begin{equation}
\widetilde{c}_{\Sigma \mathbb{P}}=R_{c^{-1}}\circ \widetilde{c}_{\Sigma
\mathbb{P}}\left( x\text{, }\xi \right) \circ R_{c}\text{.}  \label{csigP}
\end{equation}%
The pullback of $\widetilde{c}_{\Sigma \mathbb{P}}$, defined \cite%
{Tresguerres} as
\begin{equation}
\widetilde{c}_{\xi }\left( x\right) =\left( s_{M\Sigma }^{\ast }\widetilde{c}%
_{\Sigma \mathbb{P}}\right) \left( x\right) =\widetilde{c}_{\Sigma \mathbb{P}%
}\circ s_{M\Sigma }=\widetilde{c}_{\Sigma \mathbb{P}}\left( x\text{, }\xi
\right) \text{,}  \label{null-xi}
\end{equation}%
ensures the coincidence of images of sections $\widetilde{c}_{\xi }\left(
x\right) :M\rightarrow \mathbb{P}$ and $\widetilde{c}_{\Sigma \mathbb{P}%
}\left( x\text{, }\xi \right) :\Sigma \rightarrow \mathbb{P}$, respectively.
With the aid of the above results, we arrive at the useful result
\begin{equation}
\widetilde{c}_{\Sigma \mathbb{P}}\left( x\text{, }\xi \right) =\widetilde{c}%
_{M\mathbb{P}}\left( x\right) \cdot c\left( \xi \right) \text{.}
\label{csigP1}
\end{equation}

\subsection{Nonlinear Realizations and the Generalized Gauge Transformation}

The generalized gauge transformation law is obtained by comparing bundle
elements $p\in \mathbb{P}$ that differ by the left action of elements of the
principal group $G$, $L_{g\in G}$. An arbitrary element $p\in \mathbb{P}$
can be written in terms of the null section with the aid of (\ref{sMP-null}%
), (\ref{csigP}) and (\ref{csigP1}) as
\begin{equation}
p=s_{M\mathbb{P}}\left( x\right) =R_{a}\circ \widetilde{c}_{\Sigma \mathbb{P}%
}\left( x\text{, }\xi \right) \text{, }a\in H\text{.}  \label{p-initial}
\end{equation}%
Performing a gauge transformation on $p$ we obtain the orbit $\lambda \left(
p\right) $ defining a curve through $\left( x\text{, }\xi \right) $ in $%
\Sigma $%
\begin{equation}
\lambda \left( p\right) =L_{g\left( x\right) }\circ p=R_{a^{\prime }}\circ
\widetilde{c}_{\Sigma \mathbb{P}}\left( x\text{, }\xi ^{\prime }\right)
\text{; \ }g\left( x\right) \in G\text{, \ }a^{\prime }\in H\text{.}
\label{p-trans}
\end{equation}%
Comparison of (\ref{p-initial}) with (\ref{p-trans}) leads to%
\begin{equation}
L_{g\left( x\right) }\circ R_{a}\circ \widetilde{c}_{\Sigma \mathbb{P}%
}\left( x\text{, }\xi \right) =R_{a^{\prime }}\circ \widetilde{c}_{\Sigma
\mathbb{P}}\left( x\text{, }\xi ^{\prime }\right) \text{.}  \label{inter}
\end{equation}%
By virtue of the commutability \cite{Nakahara} of left and right group
translations of elements belonging to $G$, i.e. $L_{g}\circ R_{h}=R_{h}\circ
L_{g}$, (\ref{inter}) may be recast as%
\begin{equation}
L_{g\left( x\right) }\circ \widetilde{c}_{\Sigma \mathbb{P}}\left( x\text{, }%
\xi \right) =R_{h}\circ \widetilde{c}_{\Sigma \mathbb{P}}\left( x\text{, }%
\xi ^{\prime }\right) \text{.}  \label{gen-gauge}
\end{equation}%
where $R_{a^{-1}}\circ R_{a^{\prime }}\equiv R_{a^{\prime }a^{-1}}:=R_{h}$
and\ $a^{\prime }a^{-1}\equiv h\in H$. Equation (\ref{gen-gauge}) constitute
a generalized gauge transformation. Performing the pullback of (\ref%
{gen-gauge}) with respect to the section $s_{M\Sigma }$ leads to%
\begin{equation}
L_{g\left( x\right) }\circ \widetilde{c}_{\xi }\left( x\right) =R_{h\left(
\xi \text{, }g(x)\right) }\circ \widetilde{c}_{\xi ^{\prime }}\left(
x\right) \text{.}  \label{gen-gauge1}
\end{equation}%
Thus, the left action $L_{g}$ of $G$ is a map that acts on $\mathbb{P}$ and $%
\Sigma $. In particular, $L_{g}$ acting on fibers defined as orbits of the
right action describes diffeomorphisms that transforming fibers over $%
\widetilde{c}_{\xi }\left( x\right) $ into the fibers $\widetilde{c}_{\xi
^{\prime }}\left( x\right) $ of $\Sigma $ while simultaneously being
displaced along $H$ fibers via the action of $R_{h}$. Equation (\ref%
{gen-gauge1}) states that nonlinear realizations of $G$ mod $H$ is
determined by the action of an arbitrary element $g\in G$ on the quotient
space $G/H$ transforming one coset into another as
\begin{equation}
L_{g}:G/H\rightarrow G/H\text{, \ }c(\xi )\rightarrow c(\xi ^{\prime })
\end{equation}%
inducing a diffeomorphism $\xi \rightarrow \xi ^{\prime }$ on $G/H$. To
simplify the action induced by (\ref{gen-gauge1}) for calculation purposes
we proceed as follows. Departing from (\ref{null-xi}) and substituting $%
s_{M\Sigma }=R_{c}\circ \widetilde{c}_{M\mathbb{P}}$ we get%
\begin{equation}
\widetilde{c}_{\xi }\left( x\right) =\widetilde{c}_{\Sigma \mathbb{P}}\circ
R_{c}\circ \widetilde{c}_{M\Sigma }\text{.}  \label{null-xi1}
\end{equation}%
Using $\widetilde{c}_{M\mathbb{P}}\circ R_{c}=R_{c}\circ \widetilde{c}_{M%
\mathbb{P}}$, (\ref{null-xi1}) becomes $\widetilde{c}_{\xi }\left( x\right)
=R_{c}\circ \widetilde{c}_{\Sigma \mathbb{P}}\circ \widetilde{c}_{M\Sigma
}=R_{c}\circ \widetilde{c}_{M\mathbb{P}}$, where the last equality follows
from use of $\widetilde{c}_{M\mathbb{P}}=\widetilde{c}_{\Sigma \mathbb{P}%
}\circ \widetilde{c}_{M\Sigma }$. By way of analogy, we assume $\widetilde{c}%
_{\xi ^{\prime }}\left( x\right) \equiv R_{c^{\prime }}\circ \widetilde{c}_{M%
\mathbb{P}}$. Upon substitution of $\widetilde{c}_{\xi ^{\prime }}$ into (%
\ref{gen-gauge1}) we obtain%
\begin{equation}
L_{g}\circ R_{c}\circ \widetilde{c}_{M\mathbb{P}}=R_{h\left( \xi \text{, }%
g\left( x\right) \right) }\circ R_{c^{\prime }}\circ \widetilde{c}_{M\mathbb{%
P}}\text{,}
\end{equation}%
which after implementing the group actions is equivalent to,
\begin{equation}
g\cdot \widetilde{c}_{M\mathbb{P}}\cdot c=\widetilde{c}_{M\mathbb{P}}\cdot
c^{\prime }\cdot h\text{.}  \label{inter1}
\end{equation}%
Operating on (\ref{inter1}) from the left by $\widetilde{c}_{M\mathbb{P}%
}^{-1}$ and making use of $g=\widetilde{c}_{M\mathbb{P}}^{-1}g\widetilde{c}%
_{M\mathbb{P}}$, we get $\left( \widetilde{c}_{M\mathbb{P}}^{-1}\cdot g\cdot
\widetilde{c}_{M\mathbb{P}}\right) \cdot c=c^{\prime }\cdot h$ which leads
to $g\cdot c_{\xi }=c_{\xi ^{\prime }}\cdot h$, or
\begin{equation}
c^{\prime }=g\cdot c\cdot h^{-1}  \label{NLR}
\end{equation}%
in short, where $c\equiv c_{\xi }$ and $c^{\prime }\equiv c_{\xi ^{\prime }}$%
. Observe that the element $h$ is a function whose argument is the couple $%
\left( \xi \text{, }g\left( x\right) \right) $. The transformation rule (\ref%
{NLR}) is in fact the key equation to determine the nonlinear realizations
of $G$ and specifies a unique $H$-valued field $h(\xi $, $g\left( x\right) )$%
\ on $G/H$.

Consider a family of sections $\left\{ \widehat{c}\left( x\text{, }\xi
\right) \right\} $ defined \cite{TiembloTresguerres1} on $\Sigma $ by
\begin{equation}
\widehat{c}\left( x\text{, }\xi \right) :=c\circ \widetilde{c}\left( x\text{%
, }\xi \right) =c\left( \widetilde{c}\left( x\text{, }\xi \right) \right)
\text{.}  \label{null-family}
\end{equation}%
Taking $\Pi _{\mathbb{P}\Sigma }\circ R_{h}\circ $ $\widetilde{c}_{\Sigma
\mathbb{P}}=\Pi _{\mathbb{P}\Sigma }\circ \widetilde{c}_{\Sigma \mathbb{P}%
}=\left( id\right) _{\Sigma }$ into account, we can explicitly exhibit the
fact that the left action $L_{g}$ of $G$ on the null sections $\widetilde{c}%
_{\Sigma \mathbb{P}}:\mathbb{P}\rightarrow \Sigma $ induces an equivalence
relation between differing elements $\widetilde{c}_{\xi }$, $\widetilde{c}%
_{\xi ^{\prime }}\in \Sigma $ given by%
\begin{equation}
\Pi _{\mathbb{P}\Sigma }\circ L_{g}\circ \widehat{c}_{\xi }=\Pi _{\mathbb{P}%
\Sigma }\circ R_{h\left( \xi \text{, }g\left( x\right) \right) }\circ
\widehat{c}_{\xi ^{\prime }}=R_{h\left( \xi \text{, }g\left( x\right)
\right) }\circ \widetilde{c}_{\xi ^{\prime }}\text{,}
\end{equation}%
so that%
\begin{equation}
\widetilde{c}_{\xi }^{\prime }:=R_{h\left( \xi \text{, }g\left( x\right)
\right) }\circ \widetilde{c}_{\xi ^{\prime }}=L_{g}\circ \widetilde{c}_{\xi }%
\text{.}  \label{inter2}
\end{equation}%
From (\ref{inter2}) we can write%
\begin{equation}
\widetilde{c}_{\xi }\overset{L_{g}}{\longmapsto }\widetilde{c}_{\xi
}^{\prime }=R_{h\left( \xi \text{, }g\left( x\right) \right) }\circ
\widetilde{c}_{\xi ^{\prime }}\text{ }\forall h\in H\text{.}  \label{inter3}
\end{equation}%
Equation (\ref{inter3}) gives rise to a complete partition of $G/H$ into
equivalence classes $\Pi _{\mathbb{P}\Sigma }^{-1}\left( \xi \right) $ of
left cosets \cite{TiembloTresguerres1, TiembloTresguerres3}%
\begin{equation}
cH=\left\{ R_{h\left( \xi \text{, }g\left( x\right) \right) }\circ c/c\in G/H%
\text{, }\forall h\in H\right\} =\left\{ ch_{1}\text{, }ch_{2}\text{,..., }%
ch_{n}\right\} \text{,}
\end{equation}%
where $c\in (G-H)$\ plays the role of the fibers attached to each point of $%
\Sigma $. The elements $ch_{i}$\ are single representatives of each
equivalence class $R_{h\left( \xi \text{, }g\left( x\right) \right) }\circ
c=cH\in \widetilde{\pi }_{\Sigma M}^{-1}\left( \mathcal{U}\right) $. Thus,
any diffeomorphism $L_{g}\circ \widetilde{c}_{\xi }$ on $\Sigma $ together
with the $H$-valued function $h\left( \xi \text{, }g\left( x\right) \right) $
determine a unique gauge transformation $\widetilde{c}_{\xi }^{\prime
}=R_{h\left( \xi \text{, }g\left( x\right) \right) }\circ \widetilde{c}_{\xi
^{\prime }}$.\ This demonstrates that gauge transformations are those
diffeomorphisms on $\Sigma $ that map fibers over $c\left( \xi \right) $
into fibers over $c\left( \xi ^{\prime }\right) $ and simultaneously
preserves the action of $H$.

\section{Covariant Coset Field Transformations}

We now proceed to determine the transformation behavior of parameters
belonging to $G/H$. The elements of the CA and Lorentz groups are
respectively parameterized about the identity element as%
\begin{equation}
g=e^{i\epsilon ^{\alpha }\mathbf{P}_{\alpha }}e^{i\alpha ^{\mu \nu }\text{ }%
^{\dagger }\mathbf{S}_{\mu \nu }}e^{i\beta ^{\mu \nu }\mathbf{L}_{\mu \nu
}}e^{ib^{\alpha }\mathbf{\Delta }_{\alpha }}e^{i\varphi \mathbf{D}}\text{,}\
h=e^{iu^{\mu \nu }\mathbf{L}_{\mu \nu }}\text{.}
\end{equation}%
Elements of the coset space $G/H$ are coordinatized by
\begin{equation}
c=e^{-i\xi ^{\alpha }\mathbf{P}_{\alpha }}e^{ih^{\mu \nu }\text{ }^{\dagger }%
\mathbf{S}_{\mu \nu }}e^{i\zeta ^{\alpha }\mathbf{\Delta }_{\alpha
}}e^{i\phi \mathbf{D}}\text{.}
\end{equation}%
We consider transformations with infinitesimal group parameters $\epsilon
^{\alpha }$, $\alpha ^{\mu \nu }$, $\beta ^{\mu \nu }$, $b^{\alpha }$ and $%
\varphi $. The transformed coset parameters read $\xi ^{\prime \alpha }=\xi
^{\alpha }+\delta \xi ^{\alpha }$, $h^{\prime \mu \nu }=h^{\mu \nu }+\delta
h^{\mu \nu }$, $\zeta ^{\prime \alpha }=\zeta ^{\alpha }+\delta \zeta
^{\alpha }$ and $\phi ^{\prime }=\phi +\delta \phi $. Note that $u^{\mu \nu
} $ is infinitesimal. The translational coset field variations reads%
\begin{equation}
\delta \xi ^{\alpha }=-\left( \alpha _{\beta }^{\text{ \ }\alpha }+\beta
_{\beta }^{\text{ \ }\alpha }\right) \xi ^{\beta }-\epsilon ^{\alpha
}-\varphi \xi ^{\alpha }-\left[ \left\vert \xi \right\vert ^{2}b^{\alpha
}-2\left( b\cdot \xi \right) \xi ^{\alpha }\right] \text{.}
\end{equation}%
For the dilatons we get,%
\begin{equation}
\delta \phi =\varphi +2\left( b\cdot \xi \right) -\left\{ u_{\text{ }\beta
}^{\alpha }\xi ^{\beta }+\epsilon ^{\alpha }+\varphi \xi ^{\alpha }+\left[
b^{\alpha }\left\vert \xi \right\vert ^{2}-2\left( b\cdot \xi \right) \xi
^{\alpha }\right] \right\} \partial _{\alpha }\phi \text{.}
\end{equation}%
Similarly for the special conformal $4$-boosts we find,%
\begin{eqnarray}
\delta \zeta ^{\alpha } &=&u_{\text{ }\beta }^{\alpha }\zeta ^{\beta
}+b^{\alpha }-\varphi \zeta ^{\alpha }+2\left[ \left( b\cdot \xi \right)
\zeta ^{\alpha }-\left( b\cdot \zeta \right) \xi ^{\alpha }\right] + \\
&&  \notag \\
&&-\left\{ u_{\text{ }\lambda }^{\beta }\xi ^{\lambda }+\epsilon ^{\beta
}+\varphi \xi ^{\beta }+\left[ b^{\beta }\left\vert \xi \right\vert
^{2}-2\left( b\cdot \xi \right) \xi ^{\beta }\right] \right\} \partial
_{\beta }\zeta ^{\alpha }\text{.}  \notag
\end{eqnarray}%
Observe the homogeneous part of the special conformal coset parameter $\zeta
^{\alpha }$ has the same structure as that of the translational parameter $%
\xi ^{\alpha }$ (with the substitutions: $\zeta ^{\alpha }\rightarrow -\xi
^{\alpha }$ and $-\epsilon ^{\alpha }\rightarrow b^{\alpha }$). For the
shear parameters we obtain%
\begin{equation}
\delta r^{\alpha \beta }=\left( \alpha ^{\gamma \alpha }+\beta ^{\gamma
\alpha }\right) r_{\gamma }^{\text{ \ }\beta }+u_{\text{ }\gamma }^{\beta
}r^{\alpha \gamma }+2b^{[\alpha }\xi ^{\rho ]}r_{\rho }^{\text{ \ }\beta }%
\text{,}
\end{equation}%
where $r^{\alpha \beta }:=e^{h^{\alpha \beta }}$. From $\delta r^{\alpha
\beta }$\ we obtain the nonlinear Lorentz transformation%
\begin{equation}
u^{\alpha \beta }=\beta ^{\alpha \beta }+2b^{[\alpha }\xi ^{\beta ]}-\alpha
^{\mu \nu }\tanh \left\{ \frac{1}{2}\ln \left[ r_{\text{ }\mu }^{\alpha
}\left( r^{-1}\right) _{\text{ }\nu }^{\beta }\right] \right\} \text{.}
\label{NL-lorentz}
\end{equation}%
In the limit of vanishing special conformal $4$-boost, this result coincides
with that of Pinto \textit{et al.} \cite{Lopez-Pinto}. For vanishing shear,
the result of Julve \textit{et al} \cite{Julve} is obtained.

In this section, all covariant coset field transformations were determined
directly from the nonlinear transformation law (\ref{NLR}). We observe that
the translational coset parameter transforms as a coordinate under the
action of $G$. From the shear coset variation, the explicit form of the
nonlinear Lorentz-like transformation was obtained. From (\ref{NL-lorentz})
it is clear that $u^{\alpha \beta }$ contains the linear Lorentz parameter
in addition to conformal and shear contributions via the nonlinear $4$%
-boosts and symmetric $GL_{4}$ parameters.

\section{Decomposition of Connections in $\protect\pi _{\mathbb{P}M}:\mathbb{%
P}\rightarrow M$ into components in $\protect\pi _{\mathbb{P}\Sigma }:%
\mathbb{P}\rightarrow \Sigma $ and $\protect\pi _{\Sigma M}:\Sigma
\rightarrow M$}

Depending on which bundle is considered, either the total bundle $\mathbb{P}%
\rightarrow M$ or the intermediate bundles $\mathbb{P}\rightarrow \Sigma $, $%
\Sigma \rightarrow M$, we may construct corresponding Ehresmann connections
for the respective space. With respect to $M$, we have the connection form%
\begin{equation}
\omega =\widetilde{g}^{-1}\left( d+\pi _{\mathbb{P}M}^{\ast }\Omega
_{M}\right) \widetilde{g}\text{.}  \label{wPM}
\end{equation}%
The gauge potential $\Omega _{M}$ is defined in the standard manner as the
pullback of the connection $\omega $ by the null section $\widetilde{c}_{M%
\mathbb{P}}$, $\Omega _{M}=\widetilde{c}_{M\mathbb{P}}^{\ast }\omega \in
T^{\ast }\left( M\right) $. With regard to the space $\Sigma $ an
alternative form of the connection is given by%
\begin{equation}
\omega =a^{-1}\left( d+\pi _{\mathbb{P}\Sigma }^{\ast }\Gamma _{\Sigma
}\right) a\text{,}  \label{wPsig}
\end{equation}%
where the connection on $\Sigma $ reads $\Gamma _{\Sigma }=\widetilde{c}%
_{\Sigma \mathbb{P}}^{\ast }\omega $. Carrying out a similar
analysis  and evaluating the tangent vector $X\in T_{p}\left(
\Sigma \right) $ at each point $\xi $ along the curve $c_{\xi }$
on the coset space
$G/H$ that coincides with the section $\widetilde{c}_{\Sigma \mathbb{P}%
}^{\ast }$, we find the gauge transformation law%
\begin{equation}
\omega \rightarrow \omega ^{\prime }=ad_{h^{-1}}\left( d+\omega \right)
\text{.}
\end{equation}%
Comparison of (\ref{wPM}) and \ref{wPsig} leads to $\pi _{\mathbb{P}\Sigma
}^{\ast }\Gamma _{\Sigma }=c^{-1}\left( d+\pi _{\mathbb{P}M}^{\ast }\Omega
_{M}\right) c$. Taking account of $\widetilde{c}_{\Sigma \mathbb{P}}^{\ast
}\Pi _{\mathbb{P}\Sigma }^{\ast }=\left( id\right) _{T^{\ast }\left( \Sigma
\right) }$ which follows from $\Pi _{\mathbb{P}\Sigma }\circ \widetilde{c}%
_{\Sigma \mathbb{P}}=\left( id\right) _{\Sigma }$, we deduce
\begin{equation}
\Gamma _{\Sigma }=\widetilde{c}_{\Sigma \mathbb{P}}^{\ast }\left[
c^{-1}\left( d+\pi _{\mathbb{P}M}^{\ast }\Omega _{M}\right) c\right] \text{.}
\end{equation}%
By use of the family of sections pulled back to $\Sigma $ introduced in (\ref%
{null-family}) we find $\widetilde{c}_{\Sigma \mathbb{P}}^{\ast }\left(
c^{-1}dc\right) =\widehat{c}$ $^{-1}d\widehat{c}$ and\ $\widetilde{c}%
_{\Sigma \mathbb{P}}^{\ast }R_{c}^{\ast }=R_{\widehat{c}}^{\ast }\widetilde{c%
}_{\Sigma \mathbb{P}}^{\ast }$. Recalling $\widetilde{\pi }_{\mathbb{P}%
M}^{\ast }=\widetilde{\pi }_{\mathbb{P}\Sigma }^{\ast }\widetilde{\pi }%
_{\Sigma M}^{\ast }$, we get $c^{-1}\widetilde{\pi }_{\mathbb{P}M}^{\ast
}\Omega _{M}c=R_{c}^{\ast }\widetilde{\pi }_{\mathbb{P}M}^{\ast }\Omega _{M}$%
. With these results in hand, we obtain the alternative form of the
connection $\Gamma _{\Sigma }$,%
\begin{equation}
\Gamma _{\Sigma }=\widehat{c}^{-1}\left( d+\pi _{\Sigma M}^{\ast }\Omega
_{M}\right) \widehat{c}\text{.}
\end{equation}%
Completing the pullback of $\Gamma _{\Sigma }$ to $M$ by means of $%
\widetilde{c}_{M\Sigma }$ we obtain, $\Gamma _{M}=\widetilde{c}_{M\Sigma
}^{\ast }\Gamma _{\Sigma }$. By use of $\Gamma _{\Sigma }=\widetilde{c}%
_{\Sigma \mathbb{P}}^{\ast }\omega $ and (\ref{null-xi}) we find $\Gamma
_{M}=s_{M\Sigma }^{\ast }\widetilde{c}_{\Sigma \mathbb{P}}^{\ast }\omega =%
\widetilde{c}_{\xi }^{\ast }\omega $. In terms of the substitution $\widehat{%
c}\left( x\text{, }\xi \right) \rightarrow \overline{c}\left( x\right) $
where $\overline{c}\left( x\right) $ is the pullback of $\widehat{c}\left( x%
\text{, }\xi \right) $ to $M$ defined as $\overline{c}\left( x\right)
=s_{M\Sigma }^{\ast }\widehat{c}=c\left( \widetilde{c}_{\xi }\left( x\right)
\right) $, we arrive at the desired result%
\begin{equation}
\mathbf{\Gamma }\equiv \Gamma _{M}=\overline{c}^{-1}\left( d+\Omega
_{M}\right) \overline{c}\text{,}
\end{equation}%
which explicitly relates the connection $\mathbf{\Gamma }$ on $\Sigma $
pulled back to $M$ to its counterpart $\Omega _{M}$.

The gauge transformation behavior of $\mathbf{\Gamma }$\ may be determined
directly by use of (\ref{Lconn-trans}) and the transformation $\widetilde{c}%
^{\prime }=g\widetilde{c}h^{-1}$. We calculate%
\begin{equation}
\mathbf{\Gamma }^{\prime }=h\widetilde{c}^{-1}g^{-1}d\left( g\widetilde{c}%
h^{-1}\right) +h\widetilde{c}^{-1}\Omega \widetilde{c}h^{-1}+h\widetilde{c}%
^{-1}\left( dg^{-1}\right) g\widetilde{c}h^{-1}\text{.}
\end{equation}%
Observing however, that
\begin{equation}
h\widetilde{c}^{-1}g^{-1}d\left( g\widetilde{c}h^{-1}\right) =h\widetilde{c}%
^{-1}\left( g^{-1}dg\right) \widetilde{c}h^{-1}+h\widetilde{c}^{-1}d%
\widetilde{c}h^{-1}+hdh^{-1}\text{,}
\end{equation}%
we obtain%
\begin{equation}
\mathbf{\Gamma }^{\prime }=h\left[ \widetilde{c}^{-1}\left( d+\Omega \right)
\widetilde{c}\right] h^{-1}+hdh^{-1}+h\widetilde{c}^{-1}d\left(
gg^{-1}\right) \widetilde{c}h^{-1}\text{.}
\end{equation}%
Thus, we arrive at the gauge transformation law
\begin{equation}
\mathbf{\Gamma }^{\prime }=h\mathbf{\Gamma }h^{-1}+hdh^{-1}\text{.}
\label{NLR-transf}
\end{equation}%
According to the Lie algebra decomposition of $\mathfrak{g}$ into $\mathfrak{%
h}$ and $\mathfrak{c}$, the connection $\Gamma _{\Sigma }$ can be divided
into $\mathbf{\Gamma }_{H}$ defined on the subgroup $H$ and $\mathbf{\Gamma }%
_{G/H}$ defined on $G/H$. From the transformation law (\ref{NLR-transf}) it
is clear that $\mathbf{\Gamma }_{H}$ transforms inhomogeneously
\begin{equation}
\mathbf{\Gamma }_{H}^{\prime }=h\mathbf{\Gamma }_{H}h^{-1}+hdh^{-1}\text{,}
\end{equation}%
while $\Gamma _{G/H}$ transforms as a tensor
\begin{equation}
\mathbf{\Gamma }_{G/H}^{\prime }=h\mathbf{\Gamma }_{G/H}h^{-1}\text{.}
\end{equation}%
In this regard, only $\Gamma _{H}$ transforms as a true connection. We use
the gauge potential $\mathbf{\Gamma }$ to define the gauge covariant
derivative%
\begin{equation}
\mathbf{\nabla }:=\left( d+\rho \left( \mathbf{\Gamma }\right) \right)
\end{equation}%
acting on $\psi $ as $\nabla \psi =\left( d+\rho \left( \Gamma \right)
\right) \psi $ with the desired transformation property%
\begin{equation}
\left( \nabla \psi \left( c(\xi )\right) \right) ^{\prime }=\rho \left(
h(\xi \text{, }g)\right) \nabla \psi \left( c(\xi )\right) \simeq \left(
1+iu\left( \xi \text{, }g\right) \rho \left( H\right) \right) \nabla \psi
\left( c(\xi )\right)
\end{equation}%
leading to%
\begin{equation}
\delta \left( \nabla \psi \left( c(\xi )\right) \right) =iu\left( \xi \text{%
, }g\right) \rho \left( H\right) \nabla \psi \left( c(\xi )\right) \text{.}
\end{equation}

\subsection{Conform-Affine Nonlinear Gauge Potential in $\protect\pi _{%
\mathbb{P}M}:\mathbb{P\rightarrow }$ $M$}

The ordinary gauge potential defined on the total base space $M$ reads%
\begin{equation}
\Omega =-i\left( \overset{\text{T}}{\Gamma }\text{ }^{\alpha }\mathbf{P}%
_{\alpha }+\overset{\text{C}}{\Gamma }\text{ }^{\alpha }\mathbf{\Delta }%
_{\alpha }+\overset{\text{D}}{\Gamma }\mathbf{D}+\overset{\text{GL}}{\Gamma }%
\text{ }^{\alpha \beta }\text{ }^{\dagger }\mathbf{\Lambda }_{\alpha \beta
}\right) \text{.}  \label{omega}
\end{equation}%
The horizontal basis vectors that span the horizontal tangent space $\mathbb{%
H}(\mathbb{P})$ of $\pi _{\mathbb{P}M}:\mathbb{P\rightarrow }M$ are given by%
\begin{equation}
E_{i}=\widetilde{c}_{M\mathbb{P\ast }}\partial _{i}-\Omega _{i}\text{.}
\end{equation}%
The explicit form of the connections (\ref{omega}) are given by%
\begin{equation}
\omega =-i\left[ V_{M}^{\mu }\widetilde{\chi }_{\mu }^{\text{ }\nu }\mathbf{P%
}_{\nu }-i\left( i\overline{\Theta }_{\left( ^{\dagger }\mathbf{\Lambda }%
\right) }^{\alpha \beta }+\widetilde{\pi }_{\mathbb{P}M}^{\ast }\overset{%
\text{GL}}{\Gamma }\text{ }^{\alpha \beta }\right) \widetilde{\chi }_{\alpha
}^{\text{ }\nu }\widetilde{\chi }_{\beta }^{\text{ }\nu }\text{ }^{\dagger }%
\mathbf{\Lambda }_{\mu \nu }+\vartheta _{M}^{\mu }\widetilde{\beta }_{\mu }^{%
\text{ }\nu }\mathbf{\Delta }_{\nu }-i\widetilde{\pi }_{\mathbb{P}M}^{\ast
}\Phi _{M}\mathbf{D}\right]
\end{equation}%
where $\overline{\Theta }_{\left( ^{\dagger }\Lambda \right) }^{\alpha \beta
}=\overline{\Theta }_{\left( \mathbf{L}\right) }^{\alpha \beta }+\overline{%
\Theta }_{\left( \text{SY}\right) }^{\alpha \beta }$, with right invariant
Maurer-Cartan forms%
\begin{equation}
\overline{\Theta }_{\left( \mathbf{L}\right) }^{\mu \nu }=i\widetilde{\beta }%
_{\text{ \ \ \ \ }\gamma }^{[\nu |}d\widetilde{\beta }^{|\mu ]\gamma
}-2idb^{\mu }\epsilon ^{\nu }\text{ and }\overline{\Theta }_{\left( \text{SY}%
\right) }^{\mu \nu }=i\widetilde{\alpha }_{\text{ \ \ \ \ }\gamma }^{(\nu |}d%
\widetilde{\alpha }^{|\mu )\gamma }\text{.}
\end{equation}%
The linear connection $\Omega _{M}$ varies under the action of $G$ as%
\begin{equation}
\delta \Omega =\Omega ^{\prime }-\Omega =\delta \overset{\text{T}}{\Gamma }%
\text{ }^{\mu }\mathbf{P}_{\mu }+\delta \overset{\text{C}}{\Gamma }\text{ }%
^{\mu }\mathbf{\Delta }_{\mu }+\delta \overset{\text{D}}{\Gamma }\mathbf{D}%
+\delta \overset{\text{GL}}{\Gamma }\text{ }^{\beta \nu }\text{ }^{\dagger }%
\mathbf{\Lambda }_{\beta \nu }
\end{equation}%
where%
\begin{equation}
\begin{array}{c}
\delta \overset{\text{T}}{\Gamma }\text{ }^{\mu }=\text{ }^{\dagger }\overset%
{\text{GL}}{D}\epsilon ^{\mu }-\overset{\text{T}}{\Gamma }\text{ }^{\alpha
}\left( \alpha _{\alpha }^{\text{ }\mu }+\beta _{\alpha }^{\text{ }\mu
}+\varphi \delta _{\alpha }^{\text{ }\mu }\right) -\overset{\text{D}}{\Gamma
}\epsilon ^{\mu }\text{,} \\
\\
\delta \overset{\text{C}}{\Gamma }\text{ }^{\mu }=\text{ }^{\dagger }\overset%
{\text{GL}}{D}b^{\mu }-\overset{\text{C}}{\Gamma }\text{ }^{\alpha }\left(
\alpha _{\alpha }^{\text{ }\mu }+\beta _{\alpha }^{\text{ }\mu }-\varphi
\delta _{\alpha }^{\text{ }\mu }\right) +\overset{\text{D}}{\Gamma }b^{\mu }%
\text{,} \\
\\
\delta \overset{\text{GL}}{\Gamma }\text{ }^{\alpha \beta }=\text{ }%
^{\dagger }\overset{\text{GL}}{D}\left( \alpha ^{\alpha \beta }+\beta
^{\alpha \beta }\right) +\left( \overset{\text{T}}{\Gamma }\text{ }^{[\alpha
}b^{\beta ]}+\overset{\text{C}}{\Gamma }\text{ }^{[\alpha }\epsilon ^{\beta
]}\right) \text{,} \\
\\
\delta \overset{\text{D}}{\Gamma }=d\varphi +2\left( \overset{\text{C}}{%
\Gamma }\text{ }^{\alpha }\epsilon _{\alpha }-\overset{\text{T}}{\Gamma }%
\text{ }^{\alpha }b_{\alpha }\right) \text{.}%
\end{array}%
\end{equation}%
The components of $\overline{\omega }$ on $M$ are identified as spacetime
quantities and are determined from the pullback of the corresponding
(quotient space) quantities defined on $\Sigma $:%
\begin{equation}
V_{M}^{\mu }=s_{M\Sigma }^{\ast }V_{\Sigma }^{\mu }\text{,}\ \vartheta
_{M}^{\mu }=s_{M\Sigma }^{\ast }\vartheta _{\Sigma }^{\mu }\text{, }\Phi
_{M}=s_{M\Sigma }^{\ast }\Phi _{\Sigma }\ \text{and}\ \Gamma _{M}^{\mu \nu
}=s_{M\Sigma }^{\ast }\Gamma _{\Sigma }^{\mu \nu }\text{.}
\end{equation}%
In the following, we depart from the alternative form of the connection $%
\omega =a^{-1}\left( d+\Pi _{\mathbb{P}\Sigma }^{\ast }\Gamma _{\Sigma
}\right) a$, $\forall $ $a\in H$ on $\Sigma $.

\subsection{Conform-Affine Nonlinear Gauge Potential in $\protect\pi _{%
\mathbb{P}\Sigma }:\mathbb{P}\rightarrow \Sigma $}

The components of $\omega $ in $\mathbb{P}\rightarrow \Sigma $ are oriented
along the Lie algebra basis of $H$%
\begin{equation}
\overset{\mathbf{L}}{\omega }=a^{-1}\left( d+i\widetilde{\pi }_{\mathbb{P}%
\Sigma }^{\ast }\overset{\circ }{\Gamma }\text{ }^{\alpha \beta }\mathbf{L}%
_{\alpha \beta }\right) a=-i\overset{\mathbf{L}}{\omega }\text{ }^{\alpha
\beta }\mathbf{L}_{\alpha \beta }\text{,}
\end{equation}%
where%
\begin{equation}
\overset{\mathbf{L}}{\omega }\text{ }^{\alpha \beta }:=\left( i\overline{%
\Theta }_{\left( \mathbf{L}\right) }^{\rho \sigma }+\widetilde{\pi }_{%
\mathbb{P}\Sigma }^{\ast }\Gamma _{\left[ \mathbf{L}\right] }^{\rho \sigma
}\right) \widetilde{\beta }_{[\rho }^{\text{ }\alpha }\widetilde{\beta }%
_{\sigma ]}^{\text{ }\beta }\text{.}
\end{equation}

\subsection{Conform-Affine Nonlinear Gauge Potential on $\Pi _{\Sigma
M}:\Sigma \rightarrow M$}

The components of $\omega $\ in $\Pi _{\Sigma M}:\Sigma \rightarrow M$ are
oriented \cite{Tresguerres} along the Lie algebra basis of the quotient
space $G/H$ belonging to $\Sigma $%
\begin{eqnarray}
\overset{\mathbf{P}}{\omega } &=&-ia^{-1}\left( \widetilde{\pi }_{\Sigma
M}^{\ast }V_{\Sigma }^{\nu }\mathbf{P}_{\nu }\right) a=-i\overset{\mathbf{P}}%
{\omega }\text{ }^{\mu }\mathbf{P}_{\mu }\text{,} \\
&&  \notag \\
\overset{\mathbf{\Delta }}{\omega } &=&-ia^{-1}\left( \widetilde{\pi }%
_{\Sigma M}^{\ast }\vartheta _{\Sigma }^{\nu }\mathbf{\Delta }_{\nu }\right)
a=-i\overset{\mathbf{\Delta }}{\omega }\text{ }^{\mu }\mathbf{\Delta }_{\mu }%
\text{,} \\
&&  \notag \\
\overset{\mathbf{D}}{\omega } &=&-ia^{-1}\left( \widetilde{\pi }_{\Sigma
M}^{\ast }\Phi _{\Sigma }\mathbf{D}\right) a=-i\omega _{\left[ \mathbf{D}%
\right] }\mathbf{D}\text{,} \\
&&  \notag \\
\overset{\text{SY}}{\omega } &=&-ia^{-1}\left( \widetilde{\pi }_{\Sigma
M}^{\ast }\Upsilon ^{\alpha \beta }\mathbf{S}_{\alpha \beta }\right) a=-i%
\overset{\text{SY}}{\omega }\text{ }^{\alpha \beta }\mathbf{S}_{\alpha \beta
}\text{,}
\end{eqnarray}%
where%
\begin{eqnarray}
\overset{\mathbf{P}}{\omega }\text{ }^{\mu } &:&=\widetilde{\pi }_{\Sigma
M}^{\ast }V_{\Sigma }^{\nu }\widetilde{\beta }_{\nu }^{\text{ }\mu }\text{,}%
\ \overset{\mathbf{\Delta }}{\omega }\text{ }^{\mu }:=\widetilde{\pi }%
_{\Sigma M}^{\ast }\vartheta _{\Sigma }^{\nu }\widetilde{\beta }_{\nu }^{%
\text{ }\mu }\text{,} \\
&&  \notag \\
\omega _{\left[ \mathbf{D}\right] } &:&=\widetilde{\pi }_{\Sigma M}^{\ast
}\Phi _{\Sigma }\text{,}\ \overset{\text{SY}}{\omega }\text{ }^{\alpha \beta
}:=\widetilde{\pi }_{\mathbb{P}\Sigma }^{\ast }\Upsilon ^{\rho \sigma }%
\widetilde{\alpha }_{(\rho }^{\text{ }\alpha }\widetilde{\alpha }_{\sigma
)}^{\text{ }\beta }\text{.}
\end{eqnarray}%
By direct computation we obtain%
\begin{equation}
\mathbf{\Gamma }_{\Sigma }^{\text{CA}}=-i\left( V_{\Sigma }^{\mu }\mathbf{P}%
_{\mu }+i\vartheta _{\Sigma }^{\mu }\mathbf{\Delta }_{\mu }+\Phi _{\Sigma }%
\mathbf{D}+\Gamma _{\Sigma }^{\alpha \beta }\mathbf{\Lambda }_{\alpha \beta
}\right) .
\end{equation}%
The nonlinear translational and special conformal connection coefficients $%
V_{\Sigma }^{\nu }$ and $\vartheta _{\Sigma }^{\nu }$\ read%
\begin{equation}
V_{\Sigma }^{\beta }=\widetilde{\pi }_{\Sigma M}^{\ast }\left[ e^{\phi
}\left( \upsilon ^{\beta }\left( \xi \right) +r_{\text{ }\sigma }^{\alpha }%
\overset{\text{C}}{\Gamma }\text{ }^{\sigma }\mathfrak{B}_{\alpha }^{\text{ }%
\beta }\left( \xi \right) \right) \right] \text{,}  \label{NL-p1}
\end{equation}%
\begin{equation}
\vartheta _{\Sigma }^{\beta }=\widetilde{\pi }_{\Sigma M}^{\ast }\left[
e^{-\phi }\left( \upsilon ^{\beta }\left( \zeta \right) +\upsilon ^{\sigma
}\left( \xi \right) \mathfrak{B}_{\sigma }^{\text{ }\beta }\left( \zeta
\right) \right) \right] \text{,}  \label{NL-p2}
\end{equation}%
with%
\begin{equation}
\upsilon _{i}^{\beta }\left( \xi \right) :=r_{\sigma }^{\beta }\left(
\overset{\text{GL}}{^{\dagger }D_{i}}\xi ^{\sigma }+\overset{\text{D}}{%
\Gamma }_{i}\xi ^{\sigma }+\overset{\text{T}}{\Gamma }\text{ }_{i}^{\sigma
}\right) \text{, }\mathfrak{B}_{\alpha }^{\text{ }\rho }\left( \xi \right)
:=\left( \left\vert \xi \right\vert ^{2}\delta _{\alpha }^{\text{ }\rho
}-2\xi _{\alpha }\xi ^{\rho }\right) \text{.}
\end{equation}%
The nonlinear $GL_{4}$ and dilaton connections are given by%
\begin{equation}
\Gamma _{\Sigma }^{\mu \nu }=\widehat{\Gamma }\text{ }^{\mu \nu }+2\zeta
^{\lbrack \mu }\varpi ^{\nu ]}\text{,}  \label{NL-p3}
\end{equation}%
\begin{equation}
\Phi =\widetilde{\pi }_{\Sigma M}^{\ast }\left( \zeta _{\beta }\varpi
^{\beta }\right) -\frac{1}{2}d\phi \text{,}  \label{NL-p4}
\end{equation}%
with%
\begin{equation}
\widehat{\Gamma }\text{ }^{\mu \nu }:=\widetilde{\pi }_{\Sigma M}^{\ast }%
\left[ \left( r^{-1}\right) _{\;\sigma }^{\mu }\overset{\text{GL}}{\Gamma }%
\text{ }^{\sigma \beta }r_{\beta }^{\;\nu }-\left( r^{-1}\right) _{\;\sigma
}^{\mu }dr^{\sigma \nu }\right]
\end{equation}%
and%
\begin{equation}
\varpi ^{\nu }:=\upsilon ^{\nu }+r_{\text{ }\alpha }^{\nu }\overset{\text{C}}%
{\Gamma }\text{ }^{\alpha }\text{.}
\end{equation}%
The nonlinear $GL_{4}$ connection can be expanded in the $GL_{4}$ Lie
algebra according to $\Gamma ^{\alpha \beta }$ $^{\dagger }\mathbf{\Lambda }%
_{\alpha \beta }=\overset{\circ }{\Gamma }$ $^{\alpha \beta }\mathbf{L}%
_{\alpha \beta }+\Upsilon ^{\alpha \beta }$ $^{\dagger }\mathbf{S}_{\alpha
\beta }$, where%
\begin{equation}
\overset{\circ }{\Gamma }\text{ }_{\Sigma }^{\alpha \beta }:=\widehat{\Gamma
}\text{ }^{[\alpha \beta ]}+2\zeta ^{\lbrack \alpha }\varpi ^{\beta ]}\text{%
, }\Upsilon _{\Sigma }^{\alpha \beta }:=\widehat{\Gamma }\text{ }^{(\alpha
\beta )}\text{.}  \label{rest}
\end{equation}%
The symmetric $GL_{4}$ (shear) gauge fields $\Upsilon $ are distortion
fields describing the difference between the general linear connection and
the Levi-Civita connection.

We define the (group) algebra bases $e_{\nu }$ and $h_{\nu }$ dual to the
translational and special conformal 1-forms $V^{\mu }$ and $\vartheta ^{\mu
} $ as%
\begin{eqnarray}
e_{\mu } &:&=e_{\mu }^{\text{ }i}s_{M\Sigma \ast }\partial _{i}=\partial
_{\xi ^{\mu }}-e_{\mu }^{\text{ }i}\widetilde{e}_{i}\text{,} \\
&&  \notag \\
h_{\mu } &:&=h_{\mu }^{\text{ }i}s_{M\Sigma \ast }\partial _{i}=\partial
_{\zeta ^{\mu }}-h_{\mu }^{\text{ }i}\widetilde{h}_{i}\text{,}
\end{eqnarray}%
with corresponding tetrad-like components%
\begin{eqnarray}
e_{i}^{\text{ }\mu }\left( \xi \right) &=&e^{\phi }\left( \upsilon _{i}^{%
\text{ }\mu }\left( \xi \right) +r_{\text{ \ }\sigma }^{\alpha }\overset{%
\text{C}}{\Gamma }\text{ }_{i}^{\text{\ }\sigma }\mathfrak{B}_{\alpha }^{%
\text{ }\mu }\left( \xi \right) \right) \text{,}  \label{tetrad} \\
&&  \notag \\
h_{i}^{\text{ }\mu }\left( \xi \text{, }\zeta \right) &=&e^{-\phi }\left(
\upsilon _{\rho }^{\mu }\left( \zeta \right) +\upsilon _{i}^{\sigma }\left(
\xi \right) \mathfrak{B}_{\sigma }^{\text{ \ }\mu }\left( \zeta \right)
\right) \text{,}
\end{eqnarray}%
and basis vectors (on $M$)%
\begin{equation}
\widetilde{e}_{j}\left( \xi \right) =\widetilde{c}_{M\Sigma \ast }\partial
_{j}-e^{\phi }\left[ r_{\mu }^{\text{ \ }\nu }\left( \overset{\text{GL}}{%
\Gamma }\text{ }_{j\alpha }^{\text{ \ \ \ }\mu }\xi ^{\alpha }+\overset{%
\text{D}}{\Gamma }_{j}\xi ^{\mu }+\overset{\text{T}}{\Gamma }\text{ }_{j}^{%
\text{ }\mu }\right) +\overset{\text{C}}{\Gamma }\text{ }_{j}^{\text{\ }%
\sigma }r_{\text{ \ }\sigma }^{\mu }\mathfrak{B}_{\mu }^{\text{ }\nu }\left(
\xi \right) \right] \partial _{\xi ^{\nu }}
\end{equation}%
and%
\begin{equation}
\widetilde{h}_{j}\left( \xi \text{, }\zeta \right) =\widetilde{c}_{M\Sigma
\mathbb{\ast }}\partial _{j}+e^{-\phi }\left[ r_{\text{ \ }\rho }^{\mu
}\left( \overset{\text{GL}}{\Gamma }\text{ }_{j\alpha }^{\text{\ \ \ \ }\rho
}\zeta ^{\alpha }+\overset{\text{C}}{\Gamma }\text{ }_{j}^{\text{ }\rho
}\right) +r_{\text{ \ }\sigma }^{\gamma }\left( \overset{\text{GL}}{\Gamma }%
\text{ }_{j\alpha }^{\text{ \ \ \ }\sigma }\xi ^{\alpha }+\overset{\text{D}}{%
\Gamma }_{j}\xi ^{\sigma }+\overset{\text{T}}{\Gamma }\text{ }_{j}^{\text{ }%
\sigma }\right) \mathfrak{B}_{\gamma }^{\mu }\left( \zeta \right) \right]
\partial _{\zeta ^{\mu }}\text{.}
\end{equation}%
Here $\upsilon ^{\beta }\left( \zeta \right) =\upsilon ^{\beta }\left( \xi
\rightarrow \zeta \right) $, \ $\mathfrak{B}_{\alpha }^{\beta }\left( \zeta
\right) =\mathfrak{B}_{\alpha }^{\rho }\left( \xi \rightarrow \zeta \right) $%
. By definition, the basis vectors satisfy the orthogonality relations%
\begin{equation}
\left\langle V_{\Sigma }^{\mu }|\widetilde{e}_{j}\right\rangle =0\text{,\ }%
\left\langle \vartheta _{\Sigma }^{\mu }|\widetilde{h}_{j}\right\rangle =0%
\text{, }\left\langle V^{\mu }|e_{\nu }\right\rangle =\delta _{\nu }^{\mu }%
\text{,\ }\left\langle \vartheta ^{\mu }|h_{\nu }\right\rangle =\delta _{\nu
}^{\mu }\text{.}
\end{equation}%
We introduce the dilatonic and symmetric $GL_{4}$ \textit{algebra} bases%
\begin{equation}
\flat :=\partial _{\phi }-d^{i}\widetilde{d}_{i}\text{,}\ \ f_{\mu \nu
}:=\partial _{\alpha ^{\mu \nu }}-f_{\mu \nu }^{\text{ }i}\widetilde{f}_{i}
\end{equation}%
with \textit{auxiliary} \textit{soldering} components $d_{i}$ and $f_{i}^{%
\text{ }\mu \nu }$,%
\begin{eqnarray}
d_{i} &=&\zeta _{\sigma }r_{\text{ \ }\rho }^{\sigma }\left( \overset{\text{%
GL}}{^{\dagger }D_{i}}\xi ^{\rho }+\overset{\text{D}}{\Gamma }_{i}\xi ^{\rho
}+\overset{\text{T}}{\Gamma }\text{ }_{i}^{\rho }+\overset{\text{C}}{\Gamma }%
\text{ }_{i}^{\rho }\right) -\frac{1}{2}\partial _{i}\phi \text{,} \\
&&  \notag \\
f_{i}^{\text{ }\mu \nu } &=&\left( r^{-1}\right) _{\;\sigma }^{\mu }\overset{%
\text{GL}}{\Gamma }\text{ }_{i}^{\sigma \beta }r_{\beta }^{\;\nu }-\left(
r^{-1}\right) _{\;\sigma }^{\mu }\partial _{i}r^{\sigma \nu }\text{.}
\end{eqnarray}%
The \textit{coordinate} bases $\widetilde{d}_{j}$ and $\widetilde{f}_{j}$
read%
\begin{equation}
\widetilde{d}_{j}\left( \xi \text{, }\zeta \text{, }\phi \text{, }h\right) :=%
\widetilde{c}_{M\Sigma \ast }\partial _{j}-\zeta _{\sigma }r_{\text{ \ }\rho
}^{\sigma }\left( \overset{\text{GL}}{^{\dagger }\Gamma }\text{ }_{\text{ }%
j\gamma }^{\rho }\xi ^{\gamma }+\overset{\text{D}}{\Gamma }_{j}\xi ^{\rho }+%
\overset{\text{T}}{\Gamma }\text{ }_{j}^{\rho }+\overset{\text{C}}{\Gamma }%
\text{ }_{j}^{\rho }\right) \partial _{\phi }\text{,}
\end{equation}%
and%
\begin{equation}
\widetilde{f}_{j}\left( \xi \text{, }h\right) :=\widetilde{c}_{M\Sigma \ast
}\partial _{j}-\left( \left( r^{-1}\right) _{\;\ \ \ \sigma }^{(\mu |}%
\overset{\text{GL}}{\Gamma }\text{ }_{j}^{\text{ \ }\sigma \beta }r_{\beta
}^{\;\ |\nu )}-\left( r^{-1}\right) _{\;\ \ \ \sigma }^{(\mu |}\partial
_{j}r^{\sigma |\nu )}\right) \partial _{h^{\mu \nu }}\text{.}
\end{equation}%
The bases satisfy%
\begin{equation}
\left\langle \Phi |\widetilde{d}_{i}\right\rangle =0\text{, }\left\langle
\Upsilon ^{\alpha \beta }|\widetilde{f}_{i}\right\rangle =0\text{, }%
\left\langle \Phi |\flat \right\rangle =I\text{,\ }\left\langle \Upsilon
^{\alpha \beta }|f_{\mu \nu }\right\rangle =\delta _{\mu }^{\alpha }\delta
_{\nu }^{\beta }\text{.}
\end{equation}%
With the basis vectors and tetrad components in hand, we observe%
\begin{equation}
\begin{array}{c}
V_{M}^{\mu }:=dx^{i}\otimes e_{i}^{\mu }\text{,}\ \vartheta _{M}^{\mu
}:=dx^{i}\otimes h_{i}^{\mu }\text{,} \\
\\
\Phi _{M}:=dx^{i}\otimes e_{i}^{\alpha }\left\langle \Phi |e_{\alpha
}\right\rangle =dx^{i}\otimes d_{i}\text{.}%
\end{array}
\label{inter4}
\end{equation}%
The symmetric and antisymmetric $GL_{4}$ connection pulled back to $M$ is
given by%
\begin{equation}
\left.
\begin{array}{c}
\Upsilon _{M}^{\mu \nu }=dx^{i}\otimes e_{i}^{\alpha }\left\langle \Upsilon
_{\Sigma }^{\mu \nu }|e_{\alpha }\right\rangle :=dx^{i}\otimes f_{i}^{\text{
}\mu \nu }\text{,} \\
\\
\overset{\circ }{\Gamma }\text{ }_{M}^{\mu \nu }=dx^{i}\otimes e_{i}^{\alpha
}\left\langle \overset{\circ }{\Gamma }\text{ }_{\Sigma }^{\mu \nu
}|e_{\alpha }\right\rangle :=dx^{i}\otimes \overset{\circ }{\Gamma }\text{ }%
_{i}^{\mu \nu }\text{.}%
\end{array}%
\right.  \label{inter5}
\end{equation}%
With the aid of (\ref{inter4}) and (\ref{inter5}), we determine%
\begin{equation}
V_{i}^{\beta }:=e_{i}^{\text{ }\alpha }\left\langle V_{\Sigma }^{\beta
}|e_{\alpha }\right\rangle =e_{i}^{\text{ }\alpha }\delta _{\alpha }^{\beta
}=e_{i}^{\text{ }\beta }\text{, }\vartheta _{i}^{\beta }\equiv h_{i}^{\beta }%
\text{, }\Upsilon _{i}^{\mu \nu }\equiv f_{i}^{\text{ }\mu \nu }\text{, }%
\Phi _{i}\equiv d_{i}\text{.}
\end{equation}

The horizontal tangent subspace vectors in $\widetilde{\pi }_{\mathbb{P}%
\Sigma }:\mathbb{P\rightarrow }$ $\Sigma $ are given by%
\begin{equation}
\widehat{E}_{i}=\widetilde{c}_{M\mathbb{P\ast }}\widetilde{e}_{i}+i%
\widetilde{c}_{M\Sigma \mathbb{\ast }}\left\langle \overset{\circ }{\Gamma }%
\text{ }^{\alpha \beta }|\widetilde{e}_{i}\right\rangle \overset{\text{Int}}{%
\widehat{\mathfrak{R}}\text{ }_{\alpha \beta }^{\left( \mathbf{L}\right) }}%
\text{,}  \label{hor1}
\end{equation}%
\begin{equation}
\widehat{E}_{\mu }=\widetilde{c}_{\Sigma \mathbb{P\ast }}\widetilde{e}_{\mu
}+i\left\langle \overset{\circ }{\Gamma }\text{ }^{\alpha \beta }|\widetilde{%
e}_{\mu }\right\rangle \overset{\text{Int}}{\widehat{\mathfrak{R}}\text{ }%
_{\alpha \beta }^{\left( \mathbf{L}\right) }}\text{,}  \label{hor2}
\end{equation}%
and satisfy%
\begin{equation}
\left\langle \overset{\mathbf{L}}{\omega }|\widehat{E}_{j}\right\rangle
=0=\left\langle \overset{\mathbf{L}}{\omega }|\widehat{E}_{\mu
}\right\rangle \text{.}
\end{equation}%
The right invariant fundamental vector operator\emph{\ }appearing in (\ref%
{hor1})\emph{\ }or (\ref{hor2}) is given by%
\begin{equation}
\widehat{\mathfrak{R}}\text{ }_{\mu \nu }^{\left( \mathbf{L}\right)
}=i\left( \widetilde{\beta }_{[\mu |}^{\text{ \ \ \ \ }\gamma }\frac{%
\partial }{\partial \widetilde{\beta }^{|\nu ]\gamma }}+\epsilon _{\lbrack
\mu }\frac{\partial }{\partial \epsilon ^{\nu ]}}\right) \text{.}
\end{equation}%
On the other hand, the vertical tangent subspace vector in $\widetilde{\pi }%
_{\mathbb{P}\Sigma }:\mathbb{P\rightarrow }$ $\Sigma $ satisfies%
\begin{equation}
\left\langle \overset{\mathbf{L}}{\omega }|\widehat{\mathfrak{L}}\text{ }%
_{\mu \nu }^{\left( \mathbf{L}\right) }\right\rangle =\mathbf{L}_{\mu \nu
}=\left\langle \overset{\mathbf{L}}{\omega }|\widehat{\mathfrak{R}}\text{ }%
_{\mu \nu }^{\left( \mathbf{L}\right) }\right\rangle \text{,}
\end{equation}%
where%
\begin{equation}
\widehat{\mathfrak{L}}\text{ }_{\mu \nu }^{\left( \mathbf{L}\right) }=i%
\widetilde{\beta }_{\gamma \lbrack \mu |}\frac{\partial }{\partial
\widetilde{\beta }_{\gamma }^{\text{ \ }|\nu ]}}\text{, }\widehat{\mathfrak{R%
}}\text{ }_{\mu \nu }^{\left( \mathbf{L}\right) }=i\left( \widetilde{\beta }%
_{[\mu |}^{\text{ \ \ \ \ }\gamma }\frac{\partial }{\partial \widetilde{%
\beta }^{|\nu ]\gamma }}+\epsilon _{\lbrack \mu }\frac{\partial }{\partial
\epsilon ^{\nu ]}}\right) \text{.}
\end{equation}%
and $\widetilde{\beta }_{\mu }^{\text{ }\nu }:=e^{\beta _{\mu }^{\text{ }\nu
}}=\delta _{\mu }^{\text{ }\nu }+\beta _{\mu }^{\text{ }\nu }+\frac{1}{2!}%
\beta _{\mu }^{\text{ }\gamma }\beta _{\gamma }^{\text{ }\nu }+\cdot \cdot
\cdot $. The horizontal tangent subspace vectors in $\Pi _{\Sigma M}:\Sigma
\mathbb{\rightarrow }M$ are given by%
\begin{equation}
\widetilde{E}_{j}=\widetilde{c}_{\Sigma \mathbb{P\ast }}\widetilde{e}_{j}%
\text{,}\ \widetilde{H}_{j}=\widetilde{c}_{\Sigma \mathbb{P\ast }}\widetilde{%
h}_{j}\text{, }\widehat{E}\text{ }_{i}^{\left( \mathbf{D}\right) }=%
\widetilde{c}_{\Sigma \mathbb{P\ast }}\widetilde{d}_{j}\text{,}\ \overset{%
\smile }{E}_{j}=\widetilde{c}_{\Sigma \mathbb{P\ast }}\widetilde{f}_{j}\text{%
,}
\end{equation}%
and satisfy%
\begin{equation}
\left\langle \overset{\mathbf{P}}{\omega }|\widetilde{E}_{j}\right\rangle =0%
\text{, }\left\langle \overset{\mathbf{\Delta }}{\omega }|\widetilde{H}%
_{j}\right\rangle =0\text{, }\left\langle \overset{\text{SY}}{\omega }|%
\overset{\smile }{E}_{j}\right\rangle =0\text{\textbf{,\ }}\left\langle
\overset{\mathbf{D}}{\omega }|\widehat{E}\text{ }_{i}^{\left( \mathbf{D}%
\right) }\right\rangle =0\text{.}
\end{equation}%
The vertical tangent subspace vectors in $\Pi _{\Sigma M}:\Sigma \mathbb{%
\rightarrow }M$ are given by
\begin{equation}
\widetilde{E}_{\mu }=\widetilde{c}_{\Sigma \mathbb{P\ast }}\widehat{%
\mathfrak{L}}\text{ }_{\mu }^{\left( \mathbf{P}\right) }\text{,}\ \overset{%
\smile }{E}_{\alpha \beta }=\widetilde{c}_{\Sigma \mathbb{P\ast }}\widehat{%
\mathfrak{L}}\text{ }_{\alpha \beta }^{\left( \text{SY}\right) }\text{, }%
\widetilde{H}_{\mu }=\widetilde{c}_{\Sigma \mathbb{P\ast }}\widehat{%
\mathfrak{L}}\text{ }_{\mu }^{\left( \mathbf{\Delta }\right) }\text{,}\
\widehat{E}\text{ }^{\left( \mathbf{D}\right) }=\widetilde{c}_{\Sigma
\mathbb{P\ast }}\widehat{\mathfrak{L}}\text{ }^{\left( \mathbf{D}\right) }%
\text{,}  \label{vert}
\end{equation}%
and satisfy%
\begin{equation}
\left\langle \overset{\mathbf{P}}{\omega }|\widetilde{E}_{\mu }\right\rangle
=\mathbf{P}_{\mu }\text{, }\left\langle \overset{\mathbf{\Delta }}{\omega }|%
\widetilde{H}_{\mu }\right\rangle =\mathbf{\Delta }_{\mu }\text{, }%
\left\langle \overset{\text{SY}}{\omega }|\overset{\smile }{E}_{\alpha \beta
}\right\rangle =\text{ }^{\dagger }\mathbf{S}_{\alpha \beta }\text{,\ }%
\left\langle \overset{\mathbf{D}}{\omega }|\widehat{E}\text{ }^{\left(
\mathbf{D}\right) }\right\rangle =\mathbf{D}\text{.}
\end{equation}%
The left invariant fundamental vector operators\emph{\ }appearing in (\ref%
{vert})\emph{\ }are readily computed, the result being%
\begin{equation}
\begin{array}{c}
\widehat{\mathfrak{L}}\text{ }_{\mu }^{\left( \mathbf{P}\right) }=i%
\widetilde{Q}_{\text{ }\mu }^{\nu }\frac{\partial }{\partial \epsilon ^{\nu }%
}\text{, }\widehat{\mathfrak{L}}\text{ }_{\mu }^{\left( \mathbf{\Delta }%
\right) }=i\widetilde{W}_{\text{ }\mu }^{\nu }\frac{\partial }{\partial
b^{\nu }}\text{,} \\
\\
\widehat{\mathfrak{L}}\text{ }_{\alpha \beta }^{\left( \text{SY}\right) }=i%
\widetilde{\alpha }_{\gamma (\mu |}\frac{\partial }{\partial \widetilde{%
\alpha }_{\gamma }^{\text{ \ }|\nu )}}\text{, }\widehat{\mathfrak{L}}\text{ }%
^{\left( \mathbf{D}\right) }=-i\epsilon ^{\beta }\frac{\partial }{\partial
\epsilon ^{\beta }}\text{,}%
\end{array}%
\end{equation}%
where $\widetilde{\alpha }_{\mu }^{\text{ }\nu }:=e^{\alpha _{\mu }^{\text{ }%
\nu }}=\alpha _{\mu }^{\text{ }\nu }+\alpha _{\mu }^{\text{ }\nu }+\frac{1}{%
2!}\alpha _{\mu }^{\text{ }\gamma }\alpha _{\gamma }^{\text{ }\nu }+\cdot
\cdot \cdot $, $\widetilde{Q}_{\sigma }^{\text{ }\alpha }:=\left( \widetilde{%
\chi }_{\sigma }^{\text{ }\alpha }+\delta _{\sigma }^{\text{ }\alpha
}e^{\varphi }\right) $, $\widetilde{W}_{\sigma }^{\text{ }\alpha }:=\left(
\widetilde{\chi }_{\sigma }^{\text{ }\alpha }+\delta _{\sigma }^{\text{ }%
\alpha }e^{-\varphi }\right) $ satisfying $\left( \widetilde{Q}^{-1}\right)
_{\sigma }^{\text{ }\alpha }=\widetilde{Q}_{\text{ }\sigma }^{\alpha }$ and $%
\left( \widetilde{W}^{-1}\right) _{\sigma }^{\text{ }\alpha }=\widetilde{W}_{%
\text{ }\sigma }^{\alpha }$. Making use of the transformation law of the
nonlinear connection (\ref{NLR-transf}) we obtain%
\begin{equation}
\delta \Gamma =\delta V^{\alpha }\mathbf{P}_{\alpha }+\delta \vartheta
^{\alpha }\mathbf{\Delta }_{\alpha }+2\delta \Phi \mathbf{D}+\delta \mathbf{%
\Gamma }^{\alpha \beta }\text{ }^{\dagger }\mathbf{\Lambda }_{\alpha \beta }
\end{equation}%
where%
\begin{equation}
\delta V^{\nu }=u_{\alpha }^{\text{ \ }\nu }V^{\alpha }\text{,}\ \delta
\vartheta ^{\nu }=u_{\alpha }^{\text{ \ }\nu }\vartheta ^{\alpha }\text{, }%
\delta \Phi =0\text{,}\ \delta \mathbf{\Gamma }^{\alpha \beta }=\text{ }%
^{\dagger }\overset{\text{GL}}{\nabla }u^{\alpha \beta }\text{.}
\label{conn-var}
\end{equation}%
From $\delta \mathbf{\Gamma }^{\alpha \beta }=$ $^{\dagger }\overset{\text{GL%
}}{\nabla }u^{\alpha \beta }$ we observe that%
\begin{equation}
\delta \Gamma ^{\lbrack \alpha \beta ]}=\overset{\circ }{\nabla }u^{\alpha
\beta }\text{, }\delta \Upsilon _{\alpha \beta }=2u^{\rho }{}_{(\alpha
|}\Upsilon _{\rho |\beta )}\text{.}  \label{nl-pot}
\end{equation}%
According to (\ref{conn-var}), the nonlinear translational and special
conformal gauge fields transform as contravariant vector valued 1-forms
under $H$, the antisymmetric part of $\mathbf{\Gamma }^{\alpha \beta }$\
transforms inhomogeneously as a gauge potential and the nonlinear dilaton
gauge field $\Phi $ transforms as a scalar valued 1-form. From (\ref{nl-pot}%
) it is clear that the symmetric part of $\mathbf{\Gamma }^{\alpha \beta }$\
is a tensor valued 1-form. Being $4$-covectors we identify $V^{\nu }$ as
coframe fields. The connection coefficient $\overset{\circ }{\Gamma }$ $%
^{\alpha \beta }$ serves as the gravitational gauge potential. The remaining
components of $\mathbf{\Gamma }$, namely $\vartheta $, $\Upsilon $ and $\Phi
$ are dynamical fields of the theory. As will be seen in the following
subsection, the tetrad components of the coframe are used in conjunction
with the $H$-metric to induce a spacetime metric on $M$.

\section{The Induced Metric}

Since the Lorentz group $H$ is a subgroup of $G$, we inherit the invariant ($%
\delta o_{\alpha \beta }=\delta o^{\alpha \beta }=0$) (constant) metric of $%
H $, where $o^{\alpha \beta }=o_{\alpha \beta }=diag\left( -\text{, }+\text{%
, }+\text{, }+\right) $. With the aid of $o_{\alpha \beta }$ and the tetrad
components $e_{i}^{\text{ }\alpha }$ given in (\ref{tetrad}), we define the
spacetime metric
\begin{equation}
g_{ij}=e_{i}^{\text{ }\alpha }e_{j}^{\text{ }\beta }o_{\alpha \beta }\text{.}
\end{equation}%
Observing $\overset{\text{GL}}{^{\dagger }\nabla }o_{\alpha \beta
}=-2\Upsilon _{\alpha \beta }$ (where we used $do_{\alpha \beta }=0$) and
taking account of the (second) transformation property (\ref{nl-pot}), we
interpret $\Upsilon _{\alpha \beta }$ as a sort of nonmetricity, i.e. a
deformation (or distortion) gauge field that describes the difference
between the general linear connection and the Levi-Civita connection of
Riemannian geometry. In the limit of vanishing gravitational interactions $%
\overset{\text{T}}{\Gamma }$ $^{\sigma }\sim \overset{\text{C}}{\Gamma }$ $%
^{\sigma }\sim \overset{\circ }{\Gamma }$ $_{\text{ }\beta }^{\alpha }\sim
\Upsilon _{\text{ }\beta }^{\alpha }\sim \Phi \rightarrow 0$, $r_{\sigma
}^{\beta }\rightarrow \delta _{\sigma }^{\beta }$ (to first order) and $%
\overset{\text{GL}}{^{\dagger }D}\xi ^{\sigma }\rightarrow d\xi ^{\sigma }$.
Under these conditions, the coframe reduces to $V^{\beta }\rightarrow
e^{\phi }\delta _{\alpha }^{\beta }d\xi ^{\alpha }$ leading to the spacetime
metric%
\begin{equation}
g_{ij}\rightarrow e^{2\phi }\delta _{\alpha }^{\rho }\delta _{\beta
}^{\sigma }\left( \partial _{i}\xi ^{\alpha }\right) \left( \partial _{j}\xi
^{\beta }\right) o_{\rho \sigma }=e^{2\phi }\left( \partial _{i}\xi ^{\alpha
}\right) \left( \partial _{j}\xi ^{\beta }\right) o_{\alpha \beta }
\end{equation}%
characteristic of a Weyl geometry.

\section{The Cartan Structure Equations}

Using the nonlinear gauge potentials derived in  (\ref{NL-p2}),
(\ref{NL-p3}), (\ref{NL-p4}), the covariant derivative defined on
$\Sigma $ pulled back to $M$ has form
\begin{equation}
\mathbf{\nabla }:=d-iV^{\alpha }\mathbf{P}_{\alpha }-i\vartheta ^{\alpha }%
\mathbf{\Delta }_{\alpha }-2i\Phi \mathbf{D}-i\Gamma ^{\alpha \beta }\text{ }%
^{\dagger }\mathbf{\Lambda }_{\alpha \beta }.  \label{NL-covDer}
\end{equation}%
By use of (\ref{NL-covDer}) together with the relevant Lie algebra
commutators we obtain the the bundle curvature%
\begin{equation}
\mathbb{F}:=\mathbf{\nabla }\wedge \mathbf{\nabla }=-i\mathcal{T}^{\alpha }%
\mathbf{P}_{\alpha }-i\mathcal{K}^{\alpha }\mathbf{\Delta }_{\alpha }-i%
\mathcal{Z}\mathbf{D}-i\mathbb{R}_{\alpha }^{\text{ \ }\beta }\text{ }%
^{\dagger }\mathbf{\Lambda }_{\text{ \ }\beta }^{\alpha }\text{.}
\end{equation}%
The field strength components of $\mathbb{F}$\ are given by the first Cartan
structure equations. They are respectively, the projectively deformed, $%
\Upsilon $-distorted translational field strength%
\begin{equation}
\mathcal{T}^{\alpha }:=\text{ }^{\dagger }\overset{\text{GL}}{\nabla }%
V^{\alpha }+2\Phi \wedge V^{\alpha },
\end{equation}%
the projectively deformed, $\Upsilon $-distorted special conformal field
strength%
\begin{equation}
\mathcal{K}^{\alpha }:=\text{ }^{\dagger }\overset{\text{GL}}{\nabla }%
\vartheta ^{\alpha }-2\Phi \wedge \vartheta ^{\alpha },
\end{equation}%
the $\Psi $-deformed Weyl homothetic curvature 2-form (dilaton field
strength)%
\begin{equation}
\mathcal{Z}:=d\Phi +\Psi \text{,}\ \Psi =V\cdot \vartheta -\vartheta \cdot V
\end{equation}%
and the general CA curvature%
\begin{equation}
\mathbb{R}^{\alpha \beta }:=\widehat{R}\text{ }^{\alpha \beta }+\Psi
^{\alpha \beta }\text{,}
\end{equation}%
with%
\begin{equation}
\widehat{R}\text{ }^{\alpha \beta }:=\mathfrak{R}^{\alpha \beta }+\mathcal{R}%
^{\alpha \beta }\text{, \ }\Psi ^{\alpha \beta }:=V^{[\alpha }\wedge
\vartheta ^{\beta ]}\text{.}  \label{affine-curv}
\end{equation}%
Operator $^{\dagger }\overset{\text{GL}}{\nabla }$ denotes the nonlinear
covariant derivative built from volume preserving (VP) connection (i.e.
excluding $\Phi $) forms respectively. The $\Upsilon $ and $\overset{\circ }{%
\Gamma }$-affine curvatures in (\ref{affine-curv}) read%
\begin{eqnarray}
\mathfrak{R}^{\alpha \beta } &:&=\overset{\circ }{\nabla }\Upsilon ^{\alpha
\beta }+\Upsilon _{\gamma }^{\alpha }\wedge \Upsilon ^{\gamma \beta }\text{,}
\\
&&  \notag \\
\mathcal{R}^{\alpha \beta } &:&=d\overset{\circ }{\Gamma }\text{ }^{\alpha
\beta }+\overset{\circ }{\Gamma }\text{ }_{\gamma }^{\text{ }\alpha }\wedge
\overset{\circ }{\Gamma }\text{ }^{\gamma \beta }\text{,}
\end{eqnarray}%
respectively. Operator $\overset{\circ }{\nabla }$ is defined with respect
to the restricted connection $\overset{\circ }{\Gamma }$ $^{\alpha \beta }$
given in (\ref{rest}).

The field strength components of the bundle curvature have the following
group variations%
\begin{equation}
\delta \mathbb{R}_{\alpha }^{\text{ }\beta }=u_{\alpha }^{\text{ }\gamma }%
\mathbb{R}_{\text{ \ }\gamma }^{\beta }-u_{\gamma }^{\text{ }\beta }\mathbb{R%
}_{\alpha }^{\text{ }\gamma }\text{, }\delta \mathcal{Z}=0\text{, }\delta
\mathcal{T}^{\alpha }=-u_{\beta }^{\text{ }\alpha }\mathcal{T}^{\beta }\text{%
,\ }\delta \mathcal{K}^{\alpha }=-u_{\beta }^{\text{ }\alpha }\mathcal{K}%
^{\beta }\text{.}
\end{equation}%
A gauge field Lagrangian is built from polynomial combinations of the
strength $\mathbb{F}$ defined as
\begin{equation}
\mathbb{F}\left( \Gamma \left( \Omega \text{, }D\xi \right) \text{, }d\Gamma
\right) :=\nabla \wedge \nabla =d\Gamma +\Gamma \wedge \Gamma \text{.}
\end{equation}

\section{Bianchi Identities}

In what follows, the Bianchi identities (BI) play a central role. We
therefore derive them presently.

1a) The $1^{st}$ translational BI reads,%
\begin{equation}
\overset{\text{GL}}{\nabla }\mathcal{T}^{a}=\widehat{R}\text{ }_{\text{ }%
\beta }^{\alpha }\wedge V^{\beta }+\Phi \wedge T^{a}+2d\left( \Phi \wedge
V^{\alpha }\right) \text{.}
\end{equation}

1b) Similarly to the case in (1a), the $1^{st}$ conformal BIs are
respectively given by,%
\begin{equation}
\overset{\text{GL}}{\nabla }\mathcal{K}^{a}=\widehat{R}\text{ }_{\text{ }%
\beta }^{\alpha }\wedge \vartheta ^{\beta }-\Phi \wedge \mathcal{K}%
^{a}-2d\left( \Phi \wedge \vartheta ^{\alpha }\right) \text{,}
\end{equation}%
2a) The $\Upsilon $ and $\overset{\circ }{\Gamma }$-affine component of the $%
2^{nd}$ BI is given by%
\begin{equation}
^{\dagger }\overset{\text{GL}}{\nabla }\mathfrak{R}^{\alpha \beta }=2%
\mathfrak{R}_{\text{ \ \ \ }\gamma }^{(\alpha |}\Upsilon ^{\gamma |\beta )}%
\text{, }^{\dagger }\overset{\text{GL}}{\nabla }\mathcal{R}^{\alpha \beta }=0%
\text{,}
\end{equation}%
respectively. Hence, the generalized $2^{nd}$ BI is given by%
\begin{equation}
^{\dagger }\overset{\text{GL}}{\nabla }\widehat{R}\text{ }_{\text{ }\beta
}^{\alpha }=2\mathfrak{R}_{\text{ \ \ \ \ }\gamma }^{(\alpha |}\Upsilon
^{\gamma |\rho )}o_{\rho \beta }\text{.}
\end{equation}%
Since the full curvature $\mathbb{R}^{\alpha \beta }$ is proportional to $%
\Psi ^{\alpha \beta }$, it is necessary to consider%
\begin{equation}
^{\dagger }\overset{\text{GL}}{\nabla }\Psi ^{\alpha \beta }=\text{ }%
^{\dagger }\mathcal{T}^{\alpha }\wedge \vartheta ^{\beta }+V^{\alpha }\wedge
\text{ }^{\dagger }\mathcal{K}^{\beta }\text{,}
\end{equation}%
from which we conclude%
\begin{equation}
^{\dagger }\overset{\text{GL}}{\nabla }\mathbb{R}^{\alpha \beta }=2\mathfrak{%
R}_{\text{ \ \ \ }\gamma }^{(\alpha |}\Upsilon ^{\gamma |\beta )}+\text{ }%
^{\dagger }\mathcal{T}^{\alpha }\wedge \vartheta ^{\beta }+V^{\alpha }\wedge
\text{ }^{\dagger }\mathcal{K}^{\beta }.
\end{equation}%
2c) The dilatonic component of the $2^{nd}$ BI is given by%
\begin{equation}
\overset{\text{GL}}{\nabla }\mathcal{Z}=dZ+\overset{\text{GL}}{\nabla }%
\left( V\wedge \vartheta \right) =\overset{\text{GL}}{\nabla }\Psi +\Phi
\wedge \Psi \text{,}
\end{equation}%
From the definition of $\Psi $, we obtain%
\begin{equation}
\nabla \Psi =\mathcal{T}^{\alpha }\wedge \vartheta _{\alpha }+V_{\alpha
}\wedge \mathcal{K}^{\alpha }+\Phi \wedge \left( V_{\alpha }\wedge \vartheta
^{\alpha }\right) \text{.}
\end{equation}%
Defining%
\begin{equation}
\Sigma ^{\mu \nu }:=\mathbf{B}^{\mu \nu }+\Psi ^{\mu \nu }\text{,}\ \mathbf{B%
}^{\mu \nu }:=B^{\mu \nu }+\mathcal{B}^{\mu \nu }\text{, }B^{\mu \nu
}:=V^{\mu }\wedge V^{\nu }\text{, \ }\mathcal{B}^{\mu \nu }:=\vartheta ^{\mu
}\wedge \vartheta ^{\nu }\text{,}
\end{equation}%
and asserting $V^{\alpha }\wedge \vartheta _{\alpha }=0$, we find $\Sigma
_{\mu \nu }\wedge \Sigma ^{\mu \nu }=0$. Using this result,we obtain%
\begin{equation}
\nabla \Psi =\mathcal{T}^{\alpha }\wedge \vartheta _{\alpha }+V_{\alpha
}\wedge \mathcal{K}^{\alpha }\text{.}
\end{equation}

\section{Action Functional and Field Equations}

We seek an action for a local gauge theory based on the $CA\left( 3\text{, }%
1\right) $ symmetry group. We consider the $3D$ topological invariants $%
\mathbb{Y}$ of the non-Riemannian manifold of CA connections. Our objective
is the $4D$ boundary terms $\mathbb{B}$ obtained by means of exterior
differentiation of these $3D$ invariants, i.e. $\mathbb{B}=d\mathbb{Y}$. The
Lagrangian density of CA gravity is modeled after $\mathbb{B}$, with
appropriate distribution of Lie star operators so as to re-introduce the
dual frame fields. The generalized CA surface topological invariant reads%
\begin{equation}
\mathbb{Y}=-\frac{1}{2l^{2}}\left[
\begin{array}{c}
\theta _{\mathcal{A}}\left( \mathcal{A}_{a}^{\text{ }b}\wedge \widehat{R}%
\text{ }_{b}^{\text{ }a}+\frac{1}{3}\mathcal{A}_{a}^{\text{ }b}\wedge
\mathcal{A}_{b}^{\text{ }c}\wedge \mathcal{A}_{c}^{\text{ }a}\right) + \\
\\
-\theta _{\mathcal{V}}\mathcal{V}_{a}\wedge \mathbf{T}^{\alpha }+\theta
_{\Phi }\Phi \wedge \mathcal{Z}%
\end{array}%
\right] \text{,}
\end{equation}%
where $\mathbf{T}^{\alpha }:=\mathcal{T}^{\alpha }+\mathcal{K}^{\alpha }$.
The associated total CA boundary term  is given by,%
\begin{equation}
\mathbb{B}=\frac{1}{2l^{2}}\left[
\begin{array}{c}
\widehat{R}_{\beta \alpha }\wedge \mathbf{B}^{\beta \alpha }+\Sigma
^{\lbrack \beta \alpha ]}\wedge \Sigma _{\lbrack \beta \alpha ]}-\widehat{R}%
\text{ }^{\alpha \beta }\wedge \widehat{R}_{\alpha \beta }-\mathcal{Z}\wedge
\mathcal{Z}+ \\
\\
+\mathcal{K}_{\alpha }\wedge \mathcal{K}^{\alpha }+\mathcal{T}_{\alpha
}\wedge \mathcal{T}^{\alpha }-\Phi \wedge \left( V_{\alpha }\wedge \mathcal{T%
}^{\alpha }+\vartheta _{\alpha }\wedge \mathcal{K}^{\alpha }\right) + \\
\\
-\Upsilon _{\alpha \beta }\wedge \left( V^{\alpha }\wedge \mathcal{T}^{\beta
}+\vartheta ^{\alpha }\wedge \mathcal{K}^{\beta }\right) \text{.}%
\end{array}%
\right]  \label{boundary}
\end{equation}

Using the boundary term (\ref{boundary}) as a guide, we choose $[48$, $51$, $%
54$, $56,$ $66]$ an action of form%
\begin{equation}
I=\int_{\mathcal{M}}\left\{
\begin{array}{c}
d\left( \mathcal{V}^{\alpha }\wedge \mathbf{T}_{\alpha }\right) +\widehat{R}%
\text{ }^{\alpha \beta }\wedge \Sigma _{\star \alpha \beta }+\mathcal{B}%
_{\star \alpha \beta }\wedge \mathcal{B}^{\alpha \beta }+\Psi _{\star \alpha
\beta }\wedge \Psi ^{\alpha \beta }+\eta _{\star \alpha \beta }\wedge \eta
^{\alpha \beta } \\
\\
-\frac{1}{2}\left( \mathcal{R}_{\star \mu \nu }\wedge \mathcal{R}^{\mu \nu }+%
\mathcal{Z}\wedge \star \mathcal{Z}\right) +\mathcal{T}_{\star \alpha
}\wedge \mathcal{T}^{\alpha }+\mathcal{K}_{\star \alpha }\wedge \mathcal{K}%
^{\alpha }+ \\
\\
-\Phi \wedge \left( \mathcal{T}^{\star \alpha }\wedge V_{\alpha }+\mathcal{K}%
^{\star \alpha }\wedge \vartheta _{\alpha }\right) -\Upsilon _{\alpha \beta
}\wedge \left( V^{\alpha }\wedge \mathcal{T}^{\star \beta }+\vartheta
^{\alpha }\wedge \mathcal{K}^{\star \beta }\right) \text{.}%
\end{array}%
\right\}  \label{action}
\end{equation}%
Note that the action integral (\ref{action}) is invariant under Lorentz
rather than CA transformations. The Lie star $\star $ operator is defined as
$\star V_{\alpha }=\frac{1}{3!}\eta _{\alpha \beta \mu \nu }V^{\beta }\wedge
V^{\mu }\wedge V^{\nu }$.

The field equations are obtained from variation of $I$ with respect to the
independant gauge potentials. It is convenient to define the functional
derivatives%
\begin{equation}
\begin{array}{c}
\frac{\delta \mathcal{L}_{\text{gauge}}}{\delta V^{\alpha }}:=-\overset{%
\text{GL}}{\nabla }N_{\alpha }+\overset{\text{V}}{\mathfrak{T}}_{\alpha }%
\text{,} \\
\\
\frac{\delta \mathcal{L}_{\text{gauge}}}{\delta \vartheta ^{\alpha }}:=-%
\overset{\text{GL}}{\nabla }M_{\alpha }+\overset{\vartheta }{\mathfrak{T}}%
_{\alpha }\text{,} \\
\\
\mathfrak{Z}_{\alpha }^{\text{ }\beta }:=\frac{\delta \mathcal{L}_{\text{%
gauge}}}{\delta \widehat{\Gamma }\text{ }_{\text{ }\beta }^{\alpha }}=-\text{
}^{\dagger }\overset{\text{GL}}{\nabla }\widehat{M}\text{ }_{\alpha }^{\text{
}\beta }+\widehat{E}\text{ }_{\alpha }^{\text{ }\beta }\text{.}%
\end{array}%
\end{equation}%
where
\begin{equation}
\widehat{M}\text{ }_{\beta }^{\text{ }\alpha }:=-\frac{\partial \mathcal{L}_{%
\text{gauge}}}{\partial \widehat{R}\text{ }_{\text{ }\alpha }^{\beta }}\text{%
, }\widehat{E}\text{ }_{\alpha }^{\text{ }\beta }:=\frac{\partial \mathcal{L}%
_{\text{gauge}}}{\partial \widehat{\Gamma }\text{ }_{\text{ }\beta }^{\alpha
}}\text{, }\overset{\text{V}}{\mathfrak{T}}_{\alpha }:=\frac{\partial
\mathcal{L}_{\text{gauge}}}{\partial V^{\alpha }}\text{, }\overset{\vartheta
}{\mathfrak{T}}_{\alpha }:=\frac{\partial \mathcal{L}_{\text{gauge}}}{%
\partial \vartheta ^{\alpha }}\text{, }\Theta :=\frac{\partial \mathcal{L}_{%
\text{gauge}}}{\partial \Phi }\text{.}
\end{equation}%
The gauge field momenta\textit{\ }are defined by%
\begin{equation}
\begin{array}{c}
N_{\alpha }:=-\frac{\partial \mathcal{L}_{\text{gauge}}}{\partial \mathcal{T}%
^{\alpha }}\text{, }M_{\alpha }:=-\frac{\partial \mathcal{L}_{\text{gauge}}}{%
\partial \mathcal{K}^{\alpha }}\text{, }\Xi :=-\frac{\partial \mathcal{L}_{%
\text{gauge}}}{\partial \mathcal{Z}}\text{,} \\
\\
\widehat{M}_{[\alpha \beta ]}:=N_{\alpha \beta }=-o_{[\alpha |\gamma }\frac{%
\partial \mathcal{L}_{\text{gauge}}}{\partial \mathcal{R}_{\gamma }^{\text{ }%
|\beta ]}}\text{, }\widehat{M}_{(\alpha \beta )}:=M_{\alpha \beta
}=-2o_{(\alpha |\gamma }\frac{\partial \mathcal{L}_{\text{gauge}}}{\partial
\mathfrak{R}_{\gamma }^{\text{ }|\beta )}}\text{.}%
\end{array}%
\end{equation}%
Furthermore, the shear (gauge field deformation)\ and\ hypermomentum
current\ forms are given by%
\begin{equation}
\widehat{E}_{(\alpha \beta )}:=U_{\alpha \beta }=-V_{(\alpha }\wedge \left(
M_{\beta )}+N_{\beta )}\right) -M_{\alpha \beta }\text{, }\widehat{E}%
_{[\alpha \beta ]}:=E_{\alpha \beta }=-V_{[\alpha }\wedge \left( M_{\beta
]}+N_{\beta ]}\right) \text{,}
\end{equation}%
The analogue of the Einstein equations read%
\begin{equation}
G_{\alpha }+\Lambda \widehat{\eta }_{\alpha }+\text{ }^{\dagger }\overset{%
\text{GL}}{\nabla }\mathcal{T}_{\star \alpha }+\overset{\text{V}}{\mathfrak{T%
}}_{\alpha }=0\text{,}
\end{equation}%
with Einstein-like three-form%
\begin{equation}
G_{\alpha }=\left( \mathcal{R}^{\beta \gamma }+\Upsilon _{\text{ \ \ \ }\rho
}^{[\beta |}\wedge \Upsilon ^{|\gamma ]\rho }\right) \wedge \left( \eta
_{\beta \gamma \alpha }+\star \left[ B_{\beta \gamma }\wedge \vartheta
_{\alpha }\right] \right) \text{,}  \label{Einstein}
\end{equation}%
coupling constant $\Lambda $ and mixed three-form $\widehat{\eta }_{\alpha
}=\eta _{\alpha }+\star \left( \vartheta _{\alpha }\wedge V_{\beta }\right)
\wedge V^{\beta }$. Observe that $G_{\alpha }$ includes symmetric $GL_{4}$ $%
\left( \Upsilon \right) $\ as well as special conformal ($\vartheta $)
contributions. The gauge field 3-form $\overset{\text{V}}{\mathfrak{T}}%
_{\alpha }$ is given by%
\begin{eqnarray}
\overset{\text{V}}{\mathfrak{T}}_{\alpha } &=&\left\langle \mathcal{L}_{%
\text{gauge}}|e_{\alpha }\right\rangle +\left\langle \mathcal{Z}|e_{\alpha
}\right\rangle \wedge \Xi +\left\langle \mathcal{T}^{\beta }|e_{\alpha
}\right\rangle \wedge N_{\beta }+  \label{stress} \\
&&  \notag \\
&&+\left\langle \mathcal{K}^{\beta }|e_{\alpha }\right\rangle \wedge
M_{\beta }+\left\langle \mathcal{R}_{\gamma }^{\text{ }\beta }|e_{\alpha
}\right\rangle \wedge N_{\text{ }\beta }^{\gamma }+\frac{1}{2}\left\langle
\mathfrak{R}_{\gamma }^{\text{ }\beta }|e_{\alpha }\right\rangle M_{\text{ }%
\beta }^{\gamma }\text{,}  \notag
\end{eqnarray}%
We remark that to interpret (\ref{Einstein}) as the gravitational field
equation analogous to the Einstein equations, we must transform from the Lie
algebra index $\alpha $ to the spacetime basis index $k$ by contracting over
the former $\left( \alpha \right) $ with the CA tetrads $e_{k}^{\alpha }$.
\begin{eqnarray}
\overset{\text{V}}{\mathfrak{T}}_{\alpha } &=&\mathfrak{T}_{\alpha }\left[
\mathcal{T}\right] +\mathfrak{T}_{\alpha }\left[ \mathcal{K}\right] +%
\mathfrak{T}_{\alpha }\left[ \mathcal{R}\right] +\mathfrak{T}_{\alpha }\left[
Z\right] -\left\langle \mathcal{T}^{\beta }|e_{\alpha }\right\rangle \wedge
N_{\beta }-\left\langle \mathcal{K}^{\beta }|e_{\alpha }\right\rangle \wedge
M_{\beta }+ \\
&&  \notag \\
&&-\left\langle \mathcal{R}_{\gamma }^{\text{ }\beta }|e_{\alpha
}\right\rangle \wedge N_{\text{ }\beta }^{\gamma }-\left\langle \mathcal{Z}%
|e_{\alpha }\right\rangle \wedge \Xi +\Psi _{\star \alpha \beta }\wedge
\vartheta ^{\beta }+\left\langle \Sigma _{\star \gamma \beta }|e_{\alpha
}\right\rangle \wedge \widehat{R}\text{ }^{\alpha \beta }+  \notag \\
&&  \notag \\
&&+\left\langle \Upsilon ^{\gamma \beta }\wedge \left( V_{\gamma }\wedge
\mathcal{T}_{\star \beta }+\vartheta _{\gamma }\wedge \mathcal{K}_{\star
\beta }\right) |e_{\alpha }\right\rangle +\Sigma _{\star \gamma \beta
}\wedge \left\langle \widehat{R}\text{ }^{\gamma \beta }|e_{\alpha
}\right\rangle +  \notag \\
&&  \notag \\
&&\mathcal{B}_{\star \gamma \beta }\wedge \left\langle \mathcal{B}^{\gamma
\beta }|e_{\alpha }\right\rangle +\left\langle \mathcal{B}_{\star \gamma
\beta }|e_{\alpha }\right\rangle \wedge \mathcal{B}^{\gamma \beta
}+\left\langle \Psi _{\star \gamma \beta }|e_{\alpha }\right\rangle \wedge
\Psi ^{\gamma \beta }  \notag
\end{eqnarray}%
respectively, with
\begin{equation}
\begin{array}{c}
\mathfrak{T}_{\alpha }\left[ \mathcal{R}\right] =\frac{1}{2}a_{1}\left(
\mathcal{R}_{\rho \gamma }\wedge \left\langle \mathcal{R}^{\star \rho \gamma
}|e_{\alpha }\right\rangle -\left\langle \mathcal{R}_{\rho \gamma
}|e_{\alpha }\right\rangle \wedge \mathcal{R}^{\star \rho \gamma }\right)
\text{,} \\
\\
\mathfrak{T}_{\alpha }\left[ \mathcal{T}\right] =\frac{1}{2}a_{2}\left(
\mathcal{T}_{\gamma }\wedge \left\langle \mathcal{T}^{\star \gamma
}|e_{\alpha }\right\rangle -\left\langle \mathcal{T}_{\gamma }|e_{\alpha
}\right\rangle \wedge \mathcal{T}^{\star \gamma }\right) \text{,} \\
\\
\mathfrak{T}_{\alpha }\left[ \mathcal{K}\right] =\frac{1}{2}a_{3}\left(
\mathcal{K}_{\gamma }\wedge \left\langle \mathcal{K}^{\star \gamma
}|e_{\alpha }\right\rangle -\left\langle \mathcal{K}_{\gamma }|e_{\alpha
}\right\rangle \wedge \mathcal{K}^{\star \gamma }\right) \text{,} \\
\\
\mathfrak{T}_{\alpha }\left[ Z\right] =\frac{1}{2}a_{4}\left( d\Phi \wedge
\left\langle \star d\Phi |e_{\alpha }\right\rangle -\left\langle d\Phi
|e_{\alpha }\right\rangle \wedge \star d\Phi \right) \text{.}%
\end{array}%
\end{equation}%
From the variation of $I$ with respect to $\vartheta ^{\alpha }$\ we get%
\begin{equation}
\mathfrak{G}_{\alpha }+\Lambda \widehat{\omega }_{\alpha }+\text{ }^{\dagger
}\overset{\text{GL}}{\nabla }\mathcal{K}_{\star \alpha }+\overset{\vartheta }%
{\mathfrak{T}}_{\alpha }=0\text{,}
\end{equation}%
where in analogy to (\ref{Einstein}) we have%
\begin{equation}
\mathfrak{G}_{\alpha }=h_{i}^{\alpha }\left( \mathcal{R}^{\beta \gamma
}+\Upsilon _{\text{ \ \ \ }\rho }^{[\beta |}\wedge \Upsilon ^{|\gamma ]\rho
}\right) \wedge \left( \omega _{\beta \gamma \alpha }+\star \left[ \mathcal{B%
}_{\beta \gamma }\wedge V_{\alpha }\right] \right) \text{,}
\label{PDEinstein}
\end{equation}%
where $\widehat{\omega }_{\alpha }=\omega _{\alpha }+\star \left( \vartheta
_{\alpha }\wedge V_{\beta }\right) \wedge \vartheta ^{\beta }$. The quantity
$\overset{\vartheta }{\mathfrak{T}}_{i}=h_{i}^{\alpha }\overset{\vartheta }{%
\mathfrak{T}}_{\alpha }$ is similar to (\ref{stress}) but with the algebra
basis $e_{\alpha }$\ replaced by $h_{\alpha }$ and the CA tetrad components $%
e_{\text{ }i}^{\alpha }$ replaced by $h_{\text{ }i}^{\alpha }$. The two
gravitational field equations (\ref{Einstein}) and (\ref{PDEinstein}) are $%
P-\Delta $ symmetric. We may say that they exhibit $P-\Delta $ duality
symmetry invariance.

From the variational equation for $\overset{\circ }{\Gamma }$ $_{\alpha }^{%
\text{ }\beta }$ we obtain the CA gravitational analogue of the
Yang-Mills-torsion type field equation,%
\begin{equation}
\overset{\circ }{\nabla }\star \mathcal{R}_{\alpha }^{\text{ }\beta }+%
\overset{\circ }{\nabla }\star \Sigma _{\alpha }^{\text{ }\beta }+\left(
V^{\beta }\wedge \mathcal{T}_{\star \alpha }+\vartheta ^{\beta }\wedge
\mathcal{K}_{\star \alpha }\right) =0\text{.}  \label{YM}
\end{equation}%
Variation of $I$ with respect to $\Upsilon _{\alpha }^{\text{ }\beta }$
leads to%
\begin{equation}
\overset{\circ }{\nabla }\star \Sigma _{\alpha \beta }-\Upsilon _{(\alpha
|}^{\text{ \ \ \ }\gamma }\wedge \Sigma _{\star \gamma |\beta )}+V_{(\alpha
}\wedge \mathcal{T}_{\star \beta )}+\vartheta _{(\alpha }\wedge \mathcal{K}%
_{\star \beta )}=0\text{.}  \label{shear-eq}
\end{equation}%
Finally, from the variational equation for $\Phi $, the gravi-scalar field
equation is given by%
\begin{equation}
d\star d\Phi +V_{\alpha }\wedge \mathcal{T}^{\star \alpha }+\vartheta
_{\alpha }\wedge \mathcal{K}^{\star \alpha }=0\text{.}  \label{scalar}
\end{equation}

The field equations of CA gravity were obtained in this section. The
analogue of the Einstein equation, obtained from variation of $I$ with
respect to the coframe $V$, is characterized by an Einstein-like 3-form that
includes symmetric $GL_{4}$ as well as special conformal contributions.
Moreover, the field equation in (\ref{Einstein}) contains a non-trivial
torsion contribution. Performing a $P-\Delta $ transformation ( i.e. $%
V\rightarrow \vartheta $, $\mathcal{T}\rightarrow \mathcal{K}$, $%
D\rightarrow -D$) on (\ref{Einstein}) we obtain (\ref{PDEinstein}). This
result may also be obtained directly by varying $I$ with respect $\vartheta $%
. A mixed CA cosmological constant term arises in (\ref{Einstein}), (\ref{PDEinstein})%
) as a consequence of the structure of the 2-form $\mathbb{R}_{\text{ }\beta
}^{\alpha }$.

The field equation (\ref{YM}) is a Yang-Mills-like equation that represents
the generalization of the Gauss torsion-free equation $\nabla \star
B^{\alpha \beta }=0$. In our case, we considered a mixed volume form
involving both $V$ and $\vartheta $ leading to the substitution $B^{\alpha
\beta }\rightarrow \Sigma ^{\alpha \beta }$. Additionally, even in the case
of vanishing $T^{\rho }=\overset{\circ }{\nabla }V^{\rho }$, the CA torsion
depends on the dilaton potential $\Phi $ which in general is non-vanishing.
A similar argument holds for the special conformal quantity $\mathcal{K}%
^{\rho }$. Admitting the quadratic curvature term $\mathcal{R}_{\alpha
}^{\beta }\wedge \star \mathcal{R}_{\beta }^{\alpha }$ in the gauge
Lagrangian it becomes clear how we draw the analogy between (\ref{YM}) and
the Gauss equation. Equation (\ref{shear-eq}) follow from similar
considerations as (\ref{YM}), the significant differences being the lack of
a $\overset{\circ }{\nabla }\star \mathfrak{R}_{\alpha }^{\text{ }\beta }$
counterpart to $\overset{\circ }{\nabla }\star \mathcal{R}_{\alpha }^{\text{
}\beta }$ since $\star \mathfrak{R}_{\alpha }^{\text{ }\beta }=0$. Finally, (%
\ref{scalar}) involves both $\mathcal{T}^{\rho }$ and $\mathcal{K}^{\rho }$
in conjunction with a term that resembles the source-free maxwell equation
with the dilaton potential playing a similar role to the electromagnetic
vector potential.

\section{Conclusion}

In this paper a nonlinearly realized representation of the local CA group
was determined. It was found that the nonlinear Lorentz transformation law
contains contributions from the linear Lorentz parameter as well as
conformal and shear contributions via the nonlinear $4$-boosts and symmetric
$GL_{4}$ parameters. We identified the pullback of the nonlinear
translational connection coefficient to $M$ as a spacetime coframe. In this
way, the frame fields of the theory are obtained from the (nonlinear) gauge
prescription. The mixed index coframe component (tetrad) is used to convert
from Lie algebra indices into spacetime indices. The spacetime metric is a
secondary object constructed from the constant $H$ group metric and the
tetrads. The gauge fields $\overset{\circ }{\Gamma }$ $^{\alpha \beta }$ are
the analogues of the Christoffel connection coefficients of GR and serve as
the gravitational gauge potentials used to define covariant derivative
operators. The gauge fields $\vartheta $, $\Phi $, and $\Upsilon $ encode
information regarding special conformal, dilatonic and deformational degrees
of freedom of the bundle manifold. The spacetime geometry is therefore
determined by gauge field interactions.

The bundle curvature and Bianchi identities were determined. The gauge
Lagrangian density was modeled after the available boundary topological
invariants. As a consequence of this approach, no mixed field strength terms
involving different components of the total curvature arose in the action.
The analogue of the Einstein equations contains a non-trivial torsion
contribution. The Einstein-like three-form includes symmetric $GL_{4}$ as
well as special conformal contributions. A mixed translational-conformal
cosmological constant term arises due to the structure of the generalized
curvature of the manifold. We also obtain a Yang-Mills-like equation that
represents the generalization of the Gauss torsion-free equation. Variation
of $I$ with respect to $\Upsilon _{\alpha }^{\text{ }\beta }$ leads to a
constraint equation relating the $GL_{4}$ deformation gauge field to the
translational and special conformal field strengths. The gravi-scalar field
equation has non-vanishing translational and special conformal contributions.

\section{Appendix}

\subsection{Maurer-Cartan 1-forms}

For the case of matrix groups, the left invariant vector (operator)
belonging to the tangent space $\mathbb{T}(\mathbb{P})$ is defined by \cite%
{Tresguerres},%
\begin{equation}
\widehat{\mathfrak{L}}_{A}=u_{M}^{\text{ \ }L}\rho \left( \mathbf{G}%
_{A}\right) _{L}^{\text{ \ }N}\frac{\partial }{\partial u_{M}^{\text{ \ }N}}%
\text{.}
\end{equation}%
with $\left( p\widetilde{g}_{\lambda }\right) _{M}^{\text{ \ }N}=u_{M}^{%
\text{ \ }Q}\mathcal{D}_{Q}^{\text{ \ }N}$, and
$\mathcal{D}_{Q}^{\text{ \ }N}$ is the adjoint representation
matrix \cite{Lord2} for the Lie algebra basis $\mathbf{G}_{A}$.
Here $u$ is
the parameterization matrix of elements $\widetilde{g}$. For instance, if $%
\widetilde{g}=\exp (\lambda _{\text{ }B}^{A}G_{\text{ }A}^{B})$, then $u_{%
\text{ }B}^{A}:=\exp (\lambda _{\text{ }B}^{A})$. In terms of $\mathbf{G}%
_{A} $ we define the canonical $\mathfrak{g}$-valued one-form $\Theta
=g^{-1}dg=\Theta ^{A}\mathbf{G}_{A}$ $(g\in G)$ on $\mathbb{P}$, inheriting
the left invariance of $\mathbf{G}_{A}$ in terms of which it is defined,
namely $L_{g}^{\ast }\Theta |_{gp}=\Theta |_{p}$. The components of $\Theta $
read%
\begin{equation}
\Theta ^{A}=-\frac{1}{2}\left( \gamma ^{-1}\right) ^{AB}\rho \left( \mathbf{G%
}_{B}\right) _{M}^{\text{ \ \ }N}\left( u^{-1}\right) _{N}^{\text{ \ \ }%
L}du_{L}^{\text{ \ }M}\text{,}
\end{equation}%
where $\left( \gamma ^{-1}\right) ^{AB}$ is the inverse of the
Cartan-Killing metric $\gamma _{AB}$ whose anholonomic components are given
in terms of $\mathbf{G}_{A}$ as \cite{Tresguerres},
\begin{equation}
\gamma _{AB}=-2tr\left( \mathbf{G}_{A}\mathbf{G}_{B}\right) =-2f_{AM}^{\text{
\ \ \ \ \ }L}f_{BL}^{\text{ \ \ \ \ }M}\text{.}
\end{equation}%
They satisfy%
\begin{equation}
\gamma _{AB}=\mathcal{D}_{A}^{\text{ \ }C}\mathcal{D}_{B}^{\text{ \ }%
D}\gamma _{CD}\text{.}
\end{equation}%
The basis $\widehat{\mathfrak{L}}_{A}$ and one-form $\Theta $ satisfy the
duality and left invariance conditions, $\left\langle \Theta |\widehat{%
\mathfrak{L}}_{A}\right\rangle =\mathbf{G}_{A}$ and\ $L_{g\ast
}:L_{A|p}\rightarrow L_{A|gp}$. The right invariant basis vector operators
are given by%
\begin{equation}
\widehat{\mathfrak{R}}_{A}:=\rho \left( \mathbf{G}_{A}\right) _{M}^{\text{ \
\ }L}u_{L}^{\text{ \ }N}\frac{\partial }{\partial u_{M}^{\text{ \ }N}}\text{,%
}
\end{equation}%
while the canonical right invariant $\mathfrak{g}$-valued one-form $%
\overline{\Theta }=\left( dg\right) g^{-1}=\overline{\Theta }^{A}\mathbf{G}%
_{A}$, where%
\begin{equation}
\overline{\Theta }^{A}=-\frac{1}{2}\left( \gamma ^{-1}\right) ^{AB}\text{ }%
_{\rho }\left( \mathbf{G}_{B}\right) _{M}^{\text{ \ \ }N}du_{N}^{\text{ \ \ }%
L}\left( u^{-1}\right) _{L}^{\text{ \ \ }M}
\end{equation}%
satisfies $\left\langle \overline{\Theta }|\widehat{\mathfrak{R}}%
_{A}\right\rangle =\mathbf{G}_{A}$. We obtain $\Theta
^{-1}\mathbf{G}_{A}\Theta =\mathcal{D}_{A}^{\text{ \
}B}\mathbf{G}_{B}$,
where the matrix $\mathcal{D}_{A}^{\text{ }B}$ is given\ by%
\begin{equation}
\mathcal{D}_{A}^{\text{ }B}=\widehat{\mathfrak{L}}_{A}\left( \widehat{%
\mathfrak{R}}\text{ }^{-1}\right) ^{B}\text{.}
\end{equation}%
Rewriting $\mathbf{G}_{A}\Theta =\mathcal{D}_{A}^{\text{ }B}\Theta
\mathbf{G}_{B}$, differentiating with respect to
$\widetilde{g}_{\lambda }$ and taking the limit $g=\left(
id\right) _{G}$, we arrive at the commutation relations
\cite{Lord2}:
\begin{equation}
\left[ \widehat{\mathfrak{L}}_{A}\text{, }\widehat{\mathfrak{L}}_{B}\right]
=f_{AB}^{\text{ \ \ \ \ }C}\widehat{\mathfrak{L}}_{C}\text{, \ }\left[
\widehat{\mathfrak{R}}_{A}\text{, }\widehat{\mathfrak{R}}_{B}\right]
=-f_{AB}^{\text{ \ \ \ \ }C}\widehat{\mathfrak{R}}_{C}\text{, }\left[
\widehat{\mathfrak{R}}_{A}\text{, }\widehat{\mathfrak{L}}_{B}\right] =0\text{%
.}
\end{equation}%
With the aid of the BCH formula, we determine the explicit form of the
adjoint representation of the Lie algebra basis elements $ad\left(
\widetilde{g}^{-1}\right) \mathbf{G}_{A}=\mathcal{D}_{A}^{\text{ \ }B}%
\mathbf{G}_{B}$,
\begin{equation}
\mathcal{D}_{A}^{\text{ \ }B}=\left[ e^{\lambda ^{M}\rho \left( \mathbf{G}%
_{M}\right) }\right] _{A}^{\text{ \ }B}=\delta _{A}^{B}-\lambda ^{C}f_{CA}^{%
\text{ \ \ \ \ }B}+\frac{1}{2!}\lambda ^{C}f_{CA}^{\text{ \ \ \ \ }M}\lambda
^{D}f_{DM}^{\text{ \ \ \ \ \ }B}-\cdot \cdot \cdot \text{,}
\end{equation}%
where \cite{Tresguerres} use was made of $\left[ \rho \left( \mathbf{G}%
_{A}\right) \right] _{B}^{C}=-f_{AB}^{\text{ \ \ \ }C}$.

\subsection{Baker-Campbell-Hausdorff Formulas}

In the following we make extensive use of the BCH formulas%
\begin{equation}
\begin{array}{c}
e^{-A}Be^{A}=B-\frac{1}{1!}\left[ A,B\right] +\frac{1}{2!}\left[ A,\left[ A,B%
\right] \right] -\cdot \cdot \cdot \text{,} \\
\\
e^{-\chi A}de^{\chi A}=d\chi A-\frac{1}{2!}\left[ \chi A,d\chi A\right] +%
\frac{1}{3!}\left[ \chi A,\left[ \chi A,d\chi A\right] \right] -\cdot \cdot
\cdot \text{,} \\
\\
e^{i\left( h^{\mu \nu }+\delta h^{\mu \nu }\right) \text{ }^{\dagger }%
\mathbf{S}_{\mu \nu }}=e^{ih^{\mu \nu }\text{ }^{\dagger }\mathbf{S}_{\mu
\nu }}\left[ 1+ie^{-h_{\text{ }\gamma }^{\alpha }}\delta e^{h^{\gamma \beta
}}\left( ^{\dagger }\mathbf{S}_{\alpha \beta }+\mathbf{L}_{\alpha \beta
}\right) \right] \text{,} \\
\\
e^{i\left( \phi +\delta \phi \right) \mathbf{D}}=e^{i\phi \mathbf{D}}\left[
1+ie^{-h_{\text{ }\beta }^{\alpha }}\delta e^{h_{\text{ }\alpha }^{\beta }}%
\mathbf{D}\right] \text{,}%
\end{array}%
\end{equation}%
and $[70]$%
\begin{equation}
\begin{array}{c}
e^{i\xi ^{\alpha }\mathbf{P}_{\alpha }}\omega _{\alpha }^{\text{ }\beta }%
\mathbf{\Lambda }_{\text{ }\beta }^{\alpha }e^{-i\xi ^{\alpha }\mathbf{P}%
_{\alpha }}=\omega _{\alpha }^{\text{ }\beta }\mathbf{\Lambda }_{\text{ }%
\beta }^{\alpha }+\omega _{\alpha }^{\text{ }\beta }\xi ^{\alpha }\mathbf{P}%
_{\beta }\text{,} \\
\\
e^{i\Delta ^{\mu \nu }\mathbf{\Lambda }_{\mu \nu }}\kappa _{\alpha }^{\text{
}\beta }\mathbf{\Lambda }_{\text{ }\beta }^{\alpha }e^{-i\Delta ^{\mu \mu }%
\mathbf{\Lambda }_{\mu \nu }}=e^{\Delta _{\alpha }^{\text{ }\mu }}\kappa
_{\mu }^{\text{ }\nu }e^{-\Delta _{\nu }^{\text{ }\beta }}\mathbf{\Lambda }_{%
\text{ }\beta }^{\alpha }\text{,} \\
\\
e^{ih^{\mu \nu }\mathbf{S}_{\mu \nu }}\tau ^{\alpha \beta }\mathbf{L}%
_{\alpha \beta }e^{-ih^{\mu \nu }\mathbf{S}_{\mu \nu }}=e^{h_{\;\mu
}^{\alpha }}\tau ^{\mu \nu }e^{-h_{\nu }^{\;\beta }}\mathbf{\Lambda }%
_{\alpha \beta }\text{,} \\
\\
e^{ih^{\mu \nu }\mathbf{S}_{\mu \nu }}\sigma ^{\alpha \beta }\text{ }%
^{\dagger }\mathbf{S}_{\alpha \beta }e^{-ih^{\mu \nu }\mathbf{S}_{\mu \nu
}}=e^{h_{\;\mu }^{\alpha }}\sigma ^{\mu \nu }e^{-h_{\nu }^{\;\beta }}\text{ }%
^{\dagger }\mathbf{\Lambda }_{\alpha \beta }\text{,}%
\end{array}%
\end{equation}%
with $\omega _{\alpha }^{\text{ }\beta }$ $^{\dagger }\mathbf{\Lambda }%
_{\beta }^{\text{ }\alpha }=\alpha _{\alpha }^{\text{ }\beta }$ $^{\dagger }%
\mathbf{S}_{\text{ }\beta }^{\alpha }+\beta _{\alpha }^{\text{ }\beta }%
\mathbf{L}_{\text{ }\beta }^{\alpha }$. \ The components of the
stress forms
\begin{equation}
\begin{array}{c}
\alpha \wedge \star \beta =\beta \wedge \star \alpha \text{, \ }\rho \wedge
\star \sigma =\sigma \wedge \star \rho \text{,} \\
\\
\left\langle \left( \alpha \wedge \gamma \right) |v\right\rangle
=\left\langle \alpha |\nu \right\rangle \wedge \gamma +\left( -1\right)
^{p}\alpha \wedge \left\langle \gamma |\nu \right\rangle \text{,} \\
\\
\frac{\delta \left( \alpha \wedge \star \beta \right) }{\delta V}=-\delta
V^{c}\wedge \left( \left\langle \beta |e_{c}\right\rangle \wedge \star
\alpha -\left( -\right) ^{p}\alpha \wedge \left\langle \star \beta
|e_{c}\right\rangle \right) \text{,} \\
\\
\frac{\delta \left( \rho \wedge \star \sigma \right) }{\delta \vartheta }%
=-\delta \vartheta ^{c}\wedge \left( \left\langle \sigma |h_{c}\right\rangle
\wedge \star \rho -\left( -\right) ^{r}\rho \wedge \left\langle \star \sigma
|h_{c}\right\rangle \right) \text{.}%
\end{array}%
\end{equation}%
In the set of equations displayed in $\left( 4.130\right) $, $v$ is a
vector, $\alpha $ and $\beta $\ are $p$-forms that are independent of the
coframe $V$, while $\rho $ and $\sigma $\ are $r$-forms that are independent
of the special conformal coframe-like quantity $\vartheta $.

\newpage

{\LARGE Notation}

$\partial _{\mu }=\frac{\partial }{\partial x^{\mu }}$: Partial derivative
with respect to $\left\{ x_{\mu }\right\} $

$\left\{ e_{\mu }\right\} $ : Set with elements $e_{\mu }$

$\nabla _{\mu }=\partial _{\mu }+\Gamma _{\mu }$ Gauge covariant derivative
operator

$\Gamma _{\mu }$ : Gauge potential 1-form

$d$ : Exterior derivative operator

$\left\langle V|e\right\rangle $ : Inner multiplication between vector $e$
and 1-form $V$

$\left[ A\text{, }B\right] $ : Commutator of operators $A$ and $B$

$\left\{ A\text{, }B\right\} $ : Anti-commutator of operators $A$ and $B$

$\wedge $ : Exterior multiplication operator

$\rtimes $ : Semi-direct product

$\times $ : Direct product

$\times _{M}$ : Fibered product over manifold $M$

$\oplus $ : Direct sum

$\otimes $: Tensor product

$A\cup B$ : Union of $A$ and $B$

$A\cap B$ : Intersection of $A$ and $B$

$\mathbb{P}\left( M\text{, }G\text{; }\pi \right) $ : Fiber bundle with base
space $M$ and $G$-diffeomorphic fibers

$\pi _{\mathbb{P}M}:\mathbb{P}\rightarrow M$ : Canonical projection map from
$\mathbb{P}$ onto $M$

$R_{h}$, ($L_{h}$) : Right (left) group action or translation

$\widehat{\mathfrak{R}}$ ($\widehat{\mathfrak{L}}$) : Right (left) invariant
fundamental vector operators

$\Theta $ ($\overline{\Theta }$) : Right (left) invariant Maurer-Cartan
1-form

$\circ $ : Group (element) composition operator

$o_{\alpha \beta }=diag(-1$, $1$, $1$, $1)$ or $\eta _{ij}=diag(-1$, $1$, $1$%
, $1)$: Lorentz group metric

$A\left( 4\text{, }%
%TCIMACRO{\U{211d} }%
%BeginExpansion
\mathbb{R}
%EndExpansion
\right) $ : Group of affine transformations on a real 4-dimensional manifold

Diff$\left( 4\text{, }%
%TCIMACRO{\U{211d} }%
%BeginExpansion
\mathbb{R}
%EndExpansion
\right) $ : Group of diffeomorphisms on a real 4-dimensional manifold

$GL\left( 4\text{, }%
%TCIMACRO{\U{211d} }%
%BeginExpansion
\mathbb{R}
%EndExpansion
\right) $ : Group of real $4\times 4$ invertible matrices

$SO(4$, $2)$ : Special conformal group

$SO(3$, $1)$ : Lorentz group

$P(3$, $1)$ : Poincar\'{e} group

$\mathfrak{g}$ : Lie algebra of group\ $G$

$g\in G$ : Element $g$ of $G$

$\left\{ \mathcal{U}\right\} \subset M$ : Set $\mathcal{U}$ is a subset of $%
M $

$\mathbf{G}$ : Algebra generator of group $G$

$\rho \left( \mathbf{G}\right) $ : Representation of $G$-algebra

$C^{\infty }$ : Infinitely differentiable (continuous)

$^{\ast }A$ : Dual of $A$ with respect to (coordinate) basis indices

$^{\bigstar }A$ : Dual of $A$ with respect to Lie algebra indices

$\epsilon _{a_{1}...a_{n}}$ or $\varepsilon _{a_{1}...a_{n}}$ : Levi-Civita
totally skew tensor density

$\eta _{a_{1}...a_{n}}$ : Eta basis volume $n$-form density

$\sigma ^{\ast }$ : Pullback by local section $\sigma $

$L_{h\ast }$ : Differential (pushforward) map induced by $L_{h}$

$T_{\left( a_{1}...a_{n}\right) }$ : Symmetrization of indices

$T_{\left[ a_{1}...a_{n}\right] }$ : Antisymmetrization of indices

$T(M)$ : Tangent space to manifold $M$

$T^{\ast }(M)$ : Cotangent space to $M$ dual to $T(M)$

$^{\dagger }T_{\mu \nu }$ : Traceless matrix

$A^{\dagger }$ : Hermitian adjoint of $A$

$f:A\rightarrow B$ : Map $f$ taking elements $\left\{ a\right\} \in A$ to $%
\left\{ b\right\} \in B$

$h:C\hookrightarrow D$ : Inclusion map, where $C\subset D$


\begin{thebibliography}{99}

\bibitem{Utiyama} R. Utiyama, {\it Phys. Rev.} \textbf{101} (1956)
1597.

\bibitem{YangMills} C. N. Yang and R. L. Mills, {\it Phys. Rev.} \textbf{96}
(1954) 191.

\bibitem{Kibble} T.W. Kibble, Lorentz invariance and the gravitational field,
{\it J. Math. Phys.} \textbf{2} (1960) 212.

\bibitem{Cartan}  E. Cartan, {\it Ann. Ec. Norm.} {\bf 42} (1925)
17.

\bibitem{Sciama} D.W. Sciama,  On the analog between charge and spin in
general relativity, in Recent Developments in General Relativity,
Festschrift for Leopold Infeld, (1962) 415, Pergamon Press, New
York.

\bibitem{Finkelstein} R. Finkelstein, Spinor fields in spaces with torsion,
{\it Ann. Phys.} \textbf{12}, 200 (1961)

\bibitem{Hehl1} F. W. Hehl et al.,   Nonlinear spinor equation and asymmetric
connection in general relativity, {\it J. Math. Phys.} \textbf{12}
(1970) 1334.

\bibitem{Hehl2} F.W. Hehl et al.,  General relativity with spin and torsion:
foundations and prospects, {\it Rev. Mod. Phys.} \textbf{48}
(1976) 393.

\bibitem{Mansouri1} F. Mansouri et. al. Gravity as a gauge theory, {\it Phys.
Rev.} \textbf{D13} (1976) 3192.

\bibitem{Mansouri2} F. Mansouri, Conformal gravity as a gauge theory, {\it Phys.
Rev. Lett.} \textbf{42} (1979) 1021.

\bibitem{Grignani} G. Grignani et. al., Gravity and the Poincar\'{e} group,
{\it Phys. Rev.} \textbf{D45} (1992) 2719.

\bibitem{Chang} L. N. Chang et al., Geometrical approach to local gauge and
supergauge invariance: Local gauge theories and supersymmetric
strings, {\it Phys. Rev.} \textbf{D13} (1976) 235.

\bibitem{MAG} F. W. Hehl and J. D. McCrea, Bianchi Identities and the
automatic conservation of energy-momentum and angular momentum in
general-relativistic Field theories, Found. Phys. \textbf{16}
(1986) 267.

\bibitem{Inomata} A. Inomata et. al., General relativity as a limit of the de
Sitter gauge theory, {\it Phys. Rev.} \textbf{D19}  (1978) 1665.

\bibitem{CCWZ1} C. G. Callan, S. Coleman, J. Wess and B. Zumino, {\it Phys.
Rev.} \textbf{117} (1969) 2247.

\bibitem{CCWZ2} S. Coleman, J. Wess and B. Zumino, {\it Phys. Rev.} \textbf{117},
 (1969) 2239.

\bibitem{Isham} C. J. Isham, A. Salam and J. Strathdee, {\it Ann. of
Phys.} \textbf{62} (1971) 98.

\bibitem{Salam} A. Salam and J. Strathdee, {\it Phys. Rev.} \textbf{184}, 1750
(1969); {\it Phys. Rev.} \textbf{184} (1969) 1760.

\bibitem{BorisovOgievetskii} A.B. Borisov and V.I. Ogievetskii, {\it Theor. Mat.
Fiz.} \textbf{21}, 329 (1974)

\bibitem{IvanovOgievetskii} E. A. Ivanov and V. I. Ogievetskii, Gauge
theories as theories of spontaneous breakdown, Preprint of the Joint
Institute of Nuclear Research, E2-9822 (1976) 3-10

\bibitem{ChangMansouri} L. N. Chang et al., Nonlinear gauge fields and
structure of gravity and supergravity theories, {\it Phys. Rev.}
\textbf{D17} (1978) 3168.

\bibitem{StelleWest} K. S. Stelle et al., Spontaneously broken de Sitter
symmetry and the gravitational holonomy group, {\it Phys. Rev.}
\textbf{D21}  (1980) 1466.

\bibitem{IvanovNiederle1} E. A. Ivanov and J. Niederle, Gauge formulation of
gravitation theories, I. The Poincar\'{e}, de Sitter and conformal
cases, {\it Phys. Rev.} \textbf{D25} (1982) 976.

\bibitem{IvanovNiederle2} E. A. Ivanov and J. Niederle, Gauge formulation of
gravitation theories, II. The special conformal case, {\it Phys.
Rev.} \textbf{D25} (1982) 988.

\bibitem{IvanenkoSardanashvily} D. Ivanenko and G. A. Sardanashvily, {\it Phys.
Rep.} \textbf{94} (1983) 1.

\bibitem{Lord1} A. Lord and P. Goswami, Gauge theory of a group of
diffeomorphisms II: The conformal and de Sitter groups, {\it J.
Math. Phys.} \textbf{27} (1986) 3051.

\bibitem{Lord2} E. A. Lord and P. Goswami, Gauge theory of a group of
diffeomorphisms III: The fiber bundle description, {\it J. Math.
Phys.} \textbf{29}  (1987) 258.

\bibitem{NeemanRegge} Y. Ne'eman and T. Regge, {\it Riv. Nuovo Cimento} \textbf{1}
 (1978) 1.

\bibitem{NeemanSijacki} Y. Ne'eman and D. Sijacki, Gravity from Symmetry
Breakdown of a Gauge Affine Theory, The Center for Particle
Theory, University of Texas at Austin, D6-87/40 (1987).

\bibitem{Lopez-Pinto} A. Lopez-Pinto, A. Tiemblo and R. Tresguerres,
Ordinary matter in non-linear affine theories of gravitation, {\it
Class. Quant. Grav.} \textbf{12} (1995) 1503.

\bibitem{Julve} J. Julve et. al., Nonlinear 4D conformal gauge spacetime
symmetry, {\it Class. Quantum Grav.} \textbf{12} (1995) 1327.

\bibitem{TresguerresMielke} R. Tresguerres and E. W. Mielke, Gravitational
Goldstone Fields from Affine Gauge Theory, arXiv: gr-qc/0007072

\bibitem{Tresguerres} R. Tresguerres, Unified description of interactions in
terms of composite fiber bundles, {\it Phys. Rev.} \textbf{D66}
(2002) 064025.

\bibitem{TiembloTresguerres1} A. Tiemblo and R. Tresguerres, Gravitational
contribution to fermion masses, arXiv: gr-qc/0506034

\bibitem{TiembloTresguerres2} A. Tiemblo and R. Tresguerres, Recent Research
Developments in Physics, Gauge theories of gravity: the nonlinear framework,
arXiv: gr-qc/05010089

\bibitem{Schwarz} A. Schwarz, Topology for Physicists, (Springer-Verlag:
Berlin Heidelberg, 1994)

\bibitem{Nakahara} M. Nakahara, Geometry, Topology and Physics Second
Edition, Graduate Student Series in Physics, (Institute of Physics, Bristol
and Philadelphia 2003)

\bibitem{TiembloTresguerres3} A. Tiemblo and R. Tresguerres, Time evolution
in dynamical spacetimes, arXiv: gr-qc/9607066

\bibitem{Giachetta} G. Giachetta, Nonlinear realizations of the
diffeomorphism group in metric-affine gauge theory of gravity,
{\it J. Math. Phys.} \textbf{40} (1999) 939.

\bibitem{HehlCS} F. W. Hehl and W. Kopczynski, Chern-Simons terms in
metric-affine space-time: Bianchi identities as Euler-Lagrange
equations, {\it J. Math. Phys.} \textbf{32} (1991) 2169.

\bibitem{Cacciatori} S. L. Cacciatori et al., Chern-Simons formulation of
three-dimensional gravity with torsion and nonmetricity, arXiv:
hep-th/0507200

\bibitem{ChandiaZanelli} O. Chandia and J. Zanelli, Torsional Topological
Invariants (and their relevance for real life), arXiv: hep-th/9708138

\bibitem{Macias} A. Macias et. al, Torsion and Weyl covector in
metric-affine models of gravity, {\it J.Math. Phys.} \textbf{36}
(1995) 5868.

\bibitem{TresguerresCS} R. Tresguerres, Topological gravity in
three-dimensional metric-affine spacetime, {\it J. Math. Phys.}
\textbf{33} (1992) 4231.

\bibitem{TuckerWang} R. Tucker and C. Wang, Non-Riemannian Gravitational
Interactions, arXiv: gr-qc/9608055

\bibitem{Dereli} T. Dereli et al., Non-Riemannian Gravity and the Einstein
Proca System, arXiv: gr-qc/9604039

\bibitem{Scipioni} R. Scipioni, Isomorphism between Non-Riemannian Gravity
and Einstein Proca Weyl theories extended to a class of Scalar gravity
theories, arXiv: gr-qc/9905022
\end{thebibliography}
\end{document}